\documentclass{emulateapj}
\usepackage{amssymb}
\usepackage{amsmath}

\newcommand{\mbh}{\ensuremath{M_{\rm{BH}}}\,}

\newcommand{\kms}{\ifmmode {\rm km\,s}^{-1} \else km\,s$^{-1}$ \fi}

\newcommand \Hbeta {\ifmmode {\rm H}\beta \else H$\beta$\fi}
\newcommand \hb    {\ifmmode {\rm H}\beta \else H$\beta$\fi}
\newcommand \ha    {\ifmmode {\rm H}\alpha \else H$\alpha$\fi}
\newcommand  \mgii  {\ifmmode {\rm Mg}{\textsc{ii}} \else Mg\,{\sc ii}\fi}
\newcommand  \MGII  {\ifmmode {\rm Mg}\,{\sc ii}\,\lambda2798 \else Mg\,{\sc ii}\,$\lambda2798$\fi}
\newcommand  \siiv  {\ifmmode {\rm Si}\, {\sc iv}\ \else Si\,{\sc iv}\fi}
\newcommand  \SIIV  {\ifmmode {\rm Si}\,{\sc iv}\,\lambda1399 \else Si\,{\sc iv}\,$\lambda1399$\fi}
\newcommand  \civ  {\ifmmode {\rm C}\, {\sc iv}\ \else C\,{\sc iv}\fi}
\newcommand  \CIV  {\ifmmode {\rm C}\,{\sc iv}\,\lambda1549 \else C\,{\sc iv}\,$\lambda1549$\fi}
\newcommand  \NV  {\ifmmode {\rm N}\,{\sc v}\,\lambda1240 \else N\,{\sc v}\,$\lambda1240$\fi}
\newcommand  \nv  {\ifmmode {\rm N}\,{\sc v}\ \else N\,{\sc v}\fi}
\newcommand  \LyA  {\ifmmode {\rm Lyman}\,{\sc $\alpha$}\,\lambda1216 \else Lyman\,{\sc $\alpha$}\,$\lambda1216$\fi}
\newcommand  \lya {\ifmmode {\rm Lyman}\,{\sc $\alpha$}\ \else Lyman\,{\sc $\alpha$}\fi}
\newcommand  \feii     {Fe\,{\sc ii}}
\newcommand  \feiii     {Fe\,{\sc iii}}
\newcommand  \aliii  {\ifmmode {\rm Al}{\textsc{iii}} \else Al\,{\sc iii}\fi}
\newcommand  \ALIII  {\ifmmode {\rm Al}\,{\sc iii}\,\lambda1857 \else Al\,{\sc iii}\,$\lambda1857$\fi}

\newcommand  \CIII  {\ifmmode {\rm C}\,{\sc iii]}\,\lambda1909 \else C\,{\sc iii]}\,$\lambda1909$\fi}
\newcommand  \oi    {\ifmmode \left[{\rm O}\,{\textsc i}\right] \else [O\,{\sc i}]\fi}
\newcommand  \OI    {\ifmmode \left[{\rm O}\,{\textsc i}\right]\,\lambda6300 \else [O\,{\sc i}]$\,\lambda6300$ \fi}

\newcommand  \oii   {\ifmmode \left[{\rm O}\,{\textsc ii}\right] \else [O\,{\sc ii}]\fi}
\newcommand  \OII   {\ifmmode \left[{\rm O}\,{\textsc ii}\right]\,\lambda3727 \else [O\,{\sc ii}]\,$\lambda3727$ \fi}
\newcommand  \oiii  {\ifmmode \left[{\rm O}\,{\textsc iii}\right] \else [O\,{\sc iii}]\fi}
\newcommand  \OIII  {\ifmmode \left[{\rm O}\,{\textsc iii}\right]\,\lambda5007 \else [O\,{\sc iii}]\,$\lambda5007$\fi}
\newcommand  \ovi    {\ifmmode \left[{\rm O}\,{\textsc vi}\right] \else O\,{\sc vi}\fi}
\newcommand  \neiii   {\ifmmode \left[{\rm Ne}\,{\textsc iii}\right] \else [Ne\,{\sc iii}]\fi}
\newcommand  \nev   {\ifmmode \left[{\rm Ne}\,{\textsc v}\right] \else [Ne\,{\sc v}]\fi}

\shorttitle{High-$z$ BAL QSOs}
\shortauthors{Yi et al.}

\begin{document}

\title{Spectroscopy of broad absorption line quasars at $3\lesssim z \lesssim 5$ - I: evidence for quasar winds shaping broad/narrow emission line regions }
\author{ Weimin Yi\altaffilmark{1,2,8} ,Wenwen Zuo\altaffilmark{3}, Jinyi Yang\altaffilmark{4}, Feige Wang\altaffilmark{4}, John  Timlin\altaffilmark{2},  Catherine Grier\altaffilmark{4}, Xue-Bing Wu\altaffilmark{5,6}, Xiaohui Fan\altaffilmark{4},  Jin-Ming Bai\altaffilmark{1,8}  
}
\email{E-mail: ywm@ynao.ac.cn}

\altaffiltext{1}{Yunnan Observatories, Kunming, 650216, China}
\altaffiltext{2}{Department of Astronomy \& Astrophysics, The Pennsylvania State University, 525 Davey Lab, University Park, PA 16802, USA}  
\altaffiltext{3}{Shanghai Astronomical Observatory, Shanghai 200030, China}
\altaffiltext{4}{Steward Observatory, The University of Arizona, 933 North Cherry Avenue, Tucson, Arizona 85721-0065, USA}
\altaffiltext{5}{Kavli Institute for Astronomy and Astrophysics, Peking University, Beijing 100871, China}
\altaffiltext{6}{Department of Astronomy, Peking University, Yi He Yuan Lu 5, Hai Dian District, Beijing 100871, China}
\altaffiltext{7}{CAS Key Laboratory for Research in Galaxies and Cosmology, Department of Astronomy, University of Science and Technology of China, China}
\altaffiltext{8}{Key Laboratory for the Structure and Evolution of Celestial Objects, Chinese Academy of Sciences, Kunming 650216, China}



\begin{abstract}
We present an observational study of 22 broad absorption line quasars (BAL QSOs) at $3\lesssim z \lesssim5$ based on optical/near-IR spectroscopy, aiming to investigate quasar winds and their effects. 
The near-IR spectroscopy covers the \hb\ and/or \mgii\ broad emission lines (BELs) for these quasars, allowing us to estimate their central black hole (BH) masses in a robust way. 
We found that our BAL QSOs on average do not have a higher Eddington ratio than that from non-BAL QSOs matched in redshift and/or luminosity. 
In a subset consisting of seven strong BAL QSOs possessing sub-relativistic BAL outflows, we see the prevalence of large \civ-BEL blueshift ($\sim$3100 km s$^{-1}$) and weak \oiii\ emission (particularly the narrow \oiii$\lambda$5007 component), indicative of nuclear outflows affecting the narrow emission-line (NEL) regions. 
In another subset consisting of thirteen BAL QSOs having simultaneous observations of \mgii\ and \hb, we found a strong correlation between 3000~\AA\ and 5000~\AA\ monochromatic luminosity, consistent with that from non-BAL QSOs matched in redshift and luminosity; however, there is no correlation between \mgii\ and \hb\ in FWHM, likely due to nuclear outflows influencing the BEL regions. 
Our spectroscopic investigations offer strong evidence that the presence of nuclear outflows plays an important role in shaping the BEL/NEL regions of these quasars and possibly, regulating the growth of central supermassive black holes (SMBHs).  
We propose that BEL blueshift and BAL could be different manifestations of the same outflow system viewed at different sightlines and/or phases. 
\end{abstract}


\section{Introduction} \label{sec_intro}
Broad absorption line quasars (BAL QSOs, \citealp{Weymann:1991}) consist of about 20\% of the quasar population before correcting selection effects, and up to $\sim$40\% after correction (e.g., \citealp{Hewett:2003,Trump06,Gibson:2009,Allen:2011}).  BALs are unambiguous quasar winds and provide abundant diagnostics for observational studies of intrinsic outflows from low to high redshifts via \break X-ray, UV, optical, and IR spectroscopy. 
In general, most BAL QSOs are characterized by high-ionization broad absorption lines (HiBALs). 
A small fraction ($\sim$10\%, e.g. \citealp{Trump06}) of the BAL population also shows absorption troughs characterized by low-ionization species (e.g., \mgii\ and \aliii) in their spectra in addition to HiBALs, which are classified as LoBALs. 
An even rarer type of BAL QSOs show prominent \feii\ and \feiii\ absorption in addition to other low-ionization species, namely FeLoBALs.

BAL phenomena are widely interpreted by either an orientation or evolution effect. The orientation scenario appears to be  plausible for most HiBALs due to their similar continuum and emission-line properties (e.g., \citealp{Weymann:1991,Reichard03b}). However, spectropolarimetry studies suggest that BAL QSOs may not simply be normal quasars seen from an edge-on perspective (e.g., \citealp{DiPompeo10}).  
The evolution scenario has often been proposed for interpreting the phenomena of (Fe)LoBALs\footnote{Hereafter, we use (Fe)LoBAL to represent a sample including both FeLoBAL and LoBAL types.},  as they are often found from far-IR luminous and high star-formation objects (e.g., \citealp{Becker:2000,Farrah:2010}). In addition, \citet{Gallerani10} found systematic differences between BAL and non-BAL QSOs regarding observational properties that should be isotropic. Therefore, BAL QSOs might be used to investigate the early stage of quasar evolution \citep{Boroson:1992,Voit:1993,Becker:2000,Gallerani10}. In this picture,  a LoBAL QSO may be a young AGN in a short-lived transition phase between an ultra-luminous infrared galaxy (ULIRG) and a normal unobscured quasar, in which the quasar is experiencing a high-accretion process and blows off dust envelope by powerful outflows. As a consequence, such outflows may quench star formation in the host galaxy \citep{Farrah:2012,Faucher:2012}.   

\begin{table}
\center
\caption{Near-IR spectroscopic studies of BALQSOs at $z\gtrsim3$}
\begin{tabular}{ccccccc}
\hline
Reference &  Redshift  & $R$ & Num  \\
\hline
\citet{Maiolino04a} & $4.9<z<6.2$ & $\sim$50 & 4  \\ 
\citet{Gallerani10} & $3.9<z<6.2$ & 50-800 & 11 \\
\citet{Yi17} & $z=4.82$ & $\sim$350 & 1 \\
\citet{Wang18} & $z=7.02$ & $\sim$4000 & 1 \\
This work & $2.5<z<5.1$ & $\sim$2700 & 22 \\
\hline
\end{tabular}
\tablecomments{  The fourth column represents the number of BAL QSOs. $R$ is the spectral resolution. }
\label{tab_previous_bals}
\end{table}

Systematic studies of large BAL QSO samples have greatly improved our understanding of BAL structures, dynamics and intrinsic physics in both high- and low-ionization species (e.g., \citealp{Reichard03a,Gibson:2009,Schulze17,Hamann19}). 
BAL QSOs usually exhibit redder continua compared with non-BAL QSOs, which is often interpreted as stronger reddening by dust in the circumnuclear region (e.g., \citealp{Brotherton:2001,Reichard03b}). 
There is, however, an obvious deficiency of sample investigations of higher-redshift BAL QSOs based on UV/optical spectroscopy in the rest frame (see Table \ref{tab_previous_bals}), leading to a poor understanding of the BAL population in the early universe.  
The extinction curve of BAL QSOs at $z>4$, based on two dedicated studies \citep{Maiolino04a,Gallerani10}, likely deviates from the SMC extinction curve, supporting the argument of an evolutionary explanation for reddened BAL QSOs at high redshift.  Specifically, the dust production mechanism in the early universe may differ from that at low or intermediate redshift. 
However, the above two studies of high-$z$ BAL QSOs focused primarily on the investigation of internal extinction and did not  estimate BH masses in their samples (they were also limited by their low-resolution spectra, see Table \ref{tab_previous_bals}).  

At high redshift, the single-epoch spectral relation is often adopted to investigate supermassive black holes (SMBHs) residing in QSOs (e.g., \citealp{McLure2004,Vestergaard:2006}). In this relation, broad emission lines (BELs), such as H$\beta$, H$\alpha$, and \mgii, are considered to be better BH mass estimators than the \civ\ emission line. Since BAL QSOs have strong reddening and absorption, which directly affects either the BELs or the neighboring continuum, previous sample investigations of SMBHs in high-redshift QSOs are exclusively based on non-BAL QSOs (e.g., \citealp{Netzer07,Jiang07,Trakhtenbrot11,Zuo:2015,Lopez16,Coatman2017}). 
Recently, a dedicated spectroscopic study based on a sample of 22 LoBAL QSOs at $1<z<2.5$ reveals that line profiles of \hb\ and H$\alpha$ are little affected by absorption \citep{Schulze17}, indicating that they can serve as BH mass estimators for BAL QSOs. Particularly, the spectral region around the H$\alpha$ or H$\beta$ BEL is much less affected by intrinsic reddening and absorption compared to that around the \civ\ or \mgii\ BEL. Therefore, the Balmer (H$\alpha$ or H$\beta$) BELs provide unique and valuable diagnostics for the investigation of central engines powering BAL QSOs.

In this work, we present observational results from a sample including 22 BAL QSOs  at $3\lesssim z \lesssim5$ based on optical/near-IR spectroscopy. The majority of optical spectra in the sample are collected from SDSS, with a few complemented by other telescopes. We have obtained near-IR spectra for all quasars in the sample through the Telescope Access Program (TAP) from Chinese Academy of Sciences, which allows us to: (1) estimate luminosity, BH mass, and Eddington ratio for the sample; (2) systematically compare  with non-BAL quasars matched in redshift and/or luminosity, and (3)  investigate nuclear outflows and their effects at high redshift. 
Throughout this paper, a flat cosmology with $H_0$ = 70 km s$^{-1}$Mpc$^{-1}$,  $\Omega_M$ = 0.3 and  $\Omega_{\Lambda}$ = 0.7 is adopted unless stated otherwise.

\section{Sample selection and observation} 
\subsection{Sample selection}  \label{sample_selection}

We began by searching for luminous BAL QSOs from the DR12 quasar catalog \citep{Paris17} with \break AI(\civ) $\gtrsim 8000$ km s$^{-1}$ at $z>4.3$.  
To reliably measure the trough width and depth,  quasars with apparent overlapping troughs and heavy reddening imprinted on the blueward of the \civ\ emission line were excluded. Four BAL QSOs remained after this selection. We also include two more newly discovered BAL QSOs at $z>4$ that were reported by recent studies \citep{Yi15,Wang16}.

Due to the rarity of high-AI(\civ) BAL QSOs at $z>4$, we added 6 BAL QSOs showing relatively strong BAL troughs (2000$<$ AI $<$8000 km s$^{-1}$) and five comparison BAL QSOs with weak BAL troughs (AI $<$ 2000 km s$^{-1}$) at $2.5<z<4$ to our sample. Considering strong telluric-absorption windows in the near-IR wavelengths (i.e., 1.35--1.45 $\mu$m and 1.8--1.9 $\mu$m), a proper redshift cut was made to keep the \mgii\ and/or \hb\ emission lines away from  these windows. 
In total, this includes 11 BAL QSOs at $2.5<z<4$. Together with the 6 BAL QSOs at $4.3<z<5$ from the first stage and another 5 BAL QSOs at $3<z<4$ from \citet{Zuo:2015}, our final  sample consists of 22 BAL QSOs with newly and previously obtained near-IR spectra.

Basic descriptions of these BAL QSOs are tabulated in Table \ref{balqso_dr12}, where most of these measurements are retrieved  from \citet{Paris17}. The systemic redshift values are determined by the \oiii, \hb, or \mgii\ line from near-IR spectra, depending on their appearance. We visually inspected all spectra before performing quantitative measurements. 
We identified twelve LoBAL QSOs in this sample. Another three cannot be identified due to low S/N spectra and/or an  incapability of identifying low-ionization (Lo) BALs in strong telluric-absorption windows. 
The LoBAL fraction therefore is 54.5\% (12/22) or potentially higher due to unidentifiable LoBALs. We do note that this high fraction is likely caused by our preference in selecting strong BAL QSOs to construct the sample. 


\begin{table*}
\center
\caption{\civ-BAL properties of 22 BAL QSOs in the sample}
\begin{tabular}{cccccccccccc}
\hline
Name & RA & DEC & AI(CIV) & $v_{min}$ & $v_{max}$ & $v_{cen}$ & EW & $d_{\rm BAL}$ & $z$ &Type & \\
  & (J2000) & (J2000) & ($\kms$) & ($\kms$) & ($\kms$) & ($\kms$) & (\AA) &  &  & \\
\hline
014049.18$-$083942.5& 25.20492& $-$8.66181& 1760$\pm$12 & 23134 & 28496 & 25857 & 11.9$\pm$0.3 & 0.42 &3.717 & Hi \\
015048.83+004126.2 &27.70346& 0.69061& 538$\pm$4 &2913&4005 &3470 &3.4$\pm$0.2 &0.57  &3.701 &Hi \\
021646.94$-$092107.3&34.19561& $-$9.35204&2697$\pm$15 &11695&23009&13740&19.4$\pm$0.6 &0.58 &3.732 &Hi \\
074628.70+301419.0$\blacklozenge$ &116.61961&30.23863 &11060$\pm$45 &7016&30120&9600& 68.1$\pm$0.7 & 0.81 &3.173  &Lo \\
APM08279 &127.92375& 52.75486 & 1276$\pm$97 & 4125 & 10607 & 7587 & 10.0$\pm$0.9  & 0.29 & 3.911  & Hi  \\
083718.63+482806.5&129.32764&48.46848& 8068$\pm$48 & 2645&15299&8141& 48.3$\pm$0.4& 0.73 &3.64 &Lo \\
084401.95+050357.9$\blacklozenge$& 131.00813& 5.06608 & 9173$\pm$60 & 4656& 28197& 13935& 59.2$\pm$0.5 & 0.48&3.36 &Lo \\
091935.36+193834.7&139.89737&19.64297& 13189$\pm$80 & 1077& 22360& 10304& 79.2$\pm$0.6 & 0.71 &3.541 &Lo \\
103256.70+514014.5&158.23628&51.6707& 11018$\pm$66 & 1231& 27751& 9989& 69.6$\pm$0.7 & 0.62&3.925 &Lo?  \\
115023.57+281907.4&177.59825&28.31875& 788$\pm$5 & 1109& 2549& 1832& 4.92$\pm$0.3 & 0.63&3.12 &Hi  \\
120447.15+330938.7$\blacklozenge$&181.19647&33.16077& 12430$\pm$76 & 6048& 29166& 16577& 76.3$\pm$0.5 & 0.63 &3.652 &Lo  \\

121027.62+174108.9$\blacklozenge$& 182.61508 &17.68581 & 6484$\pm$87 & 14122& 29000& 20965& 41.8$\pm$0.7 & 0.5& 3.831 &Lo  \\ 
123754.82+084106.7&189.47844&8.68522& 2938$\pm$18 & 1612& 6109& 3710& 17.5$\pm$0.3 & 0.74 & 2.897 &Hi  \\
125958.72+610122.9&194.99469&61.02303& 478$\pm$3 & 3375& 4941& 4238& 3.76$\pm$0.2 & 0.39  &3.572 &Hi  \\
141546.24+112943.4& 213.94267 &11.49540 & 4896$\pm$32 & 3209 &13804 &  7533 & 31.3$\pm$0.3 & 0.56 & 2.556 &  Lo   \\
150332.17+364118.0& 225.88404& 36.68833 & 4387$\pm$20 & 3924& 28213& 5232& 32.3$\pm$0.6 & 0.76 & 3.261 & Lo  \\
\hline
\\
012247.34+121624.0$\blacklozenge$ & 20.69730 & 12.27330 & 14360$\pm$1900 & 5886 & 30479 & 14438 & 76.1$\pm$8.9 & 0.68 & 4.82  &Lo   \\
092819.28+534024.1$\blacklozenge$&142.08037&53.67337& 9711$\pm$49 & 8035& 27634& 9445& 59.7$\pm$0.8 & 0.85 &4.47 & Lo  \\
104846.63+440710.7$\blacklozenge$&162.19431&44.11966& 12258$\pm$74 & 4820& 25899& 13520& 74.3$\pm$0.6 & 0.67&4.39 & Lo  \\
133529.45+410125.9&203.87271&41.02388& 5143$\pm$32 & 16436& 26873& 20803& 32.0$\pm$0.4 & 0.59 &4.3 & Lo  \\
151035.29+514841.0& 227.64705&  51.81141& 6790$\pm$28  &  5160  &  21578 &5749 & 43.0$\pm$2 & 0.77 & 5.096 & Lo?   \\
163810.38+150058.2 &249.54325 &15.01617 & 5030$\pm$29  &  21700 & 32700 &  26850 & 31.7$\pm$3 & 0.55 & 4.84  &Lo?  \\
\hline
\end{tabular}
\tablecomments{   $v_{cen}$ is a flux-weighted centroid velocity in the \civ-BAL trough; $v_{min}$ and $v_{max}$ are the corresponding minimum and maximum velocities, respectively. EW is the equivalent width and $d_{\rm {BAL}}$ is the average depth for each \civ-BAL  BAL. Hi/Lo represent high-/low-ionization BALs, respectively. The filled diamonds indicate the seven QSOs with relatively high-S/N and $v_{min}>$ 3000 \kms, which are used to produce the composite spectra of strong BAL QSOs (see Section \ref{composite_spec}). The  Lo? symbol represents an unidentifiable LoBAL due to the lack of wavelength coverage or strong contamination from telluric absorption. }
\label{balqso_dr12}
\end{table*}

\subsection{Compilation of optical data}
Our optical spectra in the sample are primarily obtained from the DR14 SAS website \footnote{https://dr14.sdss.org/optical/spectrum/search} by matching the coordinates of the selected BAL quasars to source positions within a 3\arcsec\ tolerance. 
These spectra are corrected for calibration errors that result from atmospheric differential refraction and fiber offsets during observations (see \citealp{Margala:2016} for details). 

Additional optical spectroscopic observations were carried out during several runs from 2015 to 2018 using the Yunnan Faint Object Spectrograph and Camera mounted on the Lijiang 2.4m telescope (LJT/YFOSC, \citealt{Fan15}). 
The FWHM of the YFOSC imaging data varies mostly in the range from 1.0\arcsec\ to 2.5\arcsec; we thus chose a slit of 1.8\arcsec for the spectroscopic observations. 
Each night the Neon/Helium arc lamps were taken for wavelength calibration and a spectrophotometric standard star was observed, followed by the target at similar airmass and position. 
All the calibration data, including bias, sky-flat fields, and internal lamp flats were obtained at the beginning/end of each observing night.  
The data reduction was performed by IRAF packages following standard reduction steps for bias/sky subtractions, flat-fielding correction and flux calibration.

\subsection{Near-IR spectroscopic observations and data reduction}
The near-IR spectroscopic observations of all objects tabulated in Table \ref{balqso_dr12} were carried out with the TripleSpec spectrograph \citep{Wilson:2004} at the Palomar Hale 200 inch telescope (P200/TripleSpec, \citealp{Wilson:2004}). 
TripleSpec provides a wide wavelength coverage from 0.95 $\mu$m to 2.46 $\mu$m at an average spectral resolution of $R\sim2700$, allowing simultaneous observations in the JHK bands. High Fowler depth ($6\sim10$) is set for faint objects to optimize the readout noise after each single exposure. A slit width of 1\arcsec\  and the ABBA dither pattern along the slit were chosen to improve the sky subtraction for all targets  throughout these observational runs. 
Total exposure times varied between 24 and 60~minutes, depending on the apparent magnitudes and weather conditions. During each night, several telluric A0V standard stars were observed at a similar airmass compared to that  of the target.

The data reduction for the spectroscopic data from TripleSpec is performed using the modified IDL-based Spextool3 package
\citep{Cushing:2004}, as described in detail from \citet{Zuo:2015}. The standard process includes sky background subtraction, flat field correction,  wavelength identification and telluric correction. Based on the reduced near-IR spectra, we determine systemic redshift using the \oiii, or \hb, or \mgii\ emission line, ordered by a descending priority when available. 

\subsection{Notes on individual objects}
J0122+1216 was discovered as a quasar using the Lijiang 2.4-m Telescope \citep{Yi15} in optical spectroscopy; later, it was further identified as a LoBAL QSO at $z=4.82$ using near-IR spectroscopy \citep{Yi17}. Recently, this LoBAL QSO was investigated  by \citet{Jeon17}, where their measurements of the luminosity and BH mass are consistent with those from \citet{Yi17}. As the quality of near-IR spectra obtained by the Magellan telescope (see \citealp{Jeon17}) is much higher than that from \citet{Yi17}, we use their near-IR spectra (in private communication) for J0122+1216 in this work. 

APM08279 is one of the brightest quasars; its brightness is due to gravitational lensing \citep{Irwin98}.  J1415+1129 is another lensed quasar showing a ``cloverleaf'' structure \citep{Magain88}.  
Since previous studies reported a wide range of possible magnification factors (4 $\sim$ 100) for this quasar, for simplicity, we choose to use the  bolometric luminosity assuming that it is equal to the Eddington luminosity limit. 

\section{ Spectroscopic measurements and analyses}  \label{sec:methods}
We use several different methods to obtain more reliable measurements based on optical/near-IR spectroscopic/photometric data. In particular, we focus on redshift determination, similar spectral shape matching at close redshifts, continuum fitting, correction of absolute spectral flux, and emission line fitting. The relevant measurements are tabulated in Table \ref{balqso_dr12} and Table \ref{tab:prop}. 

\subsection{Measurements from optical spectra} \label{blueshift_determine}

 \begin{figure} 
  \center{}
      \includegraphics[width=8.5cm, height=9cm, angle=0]{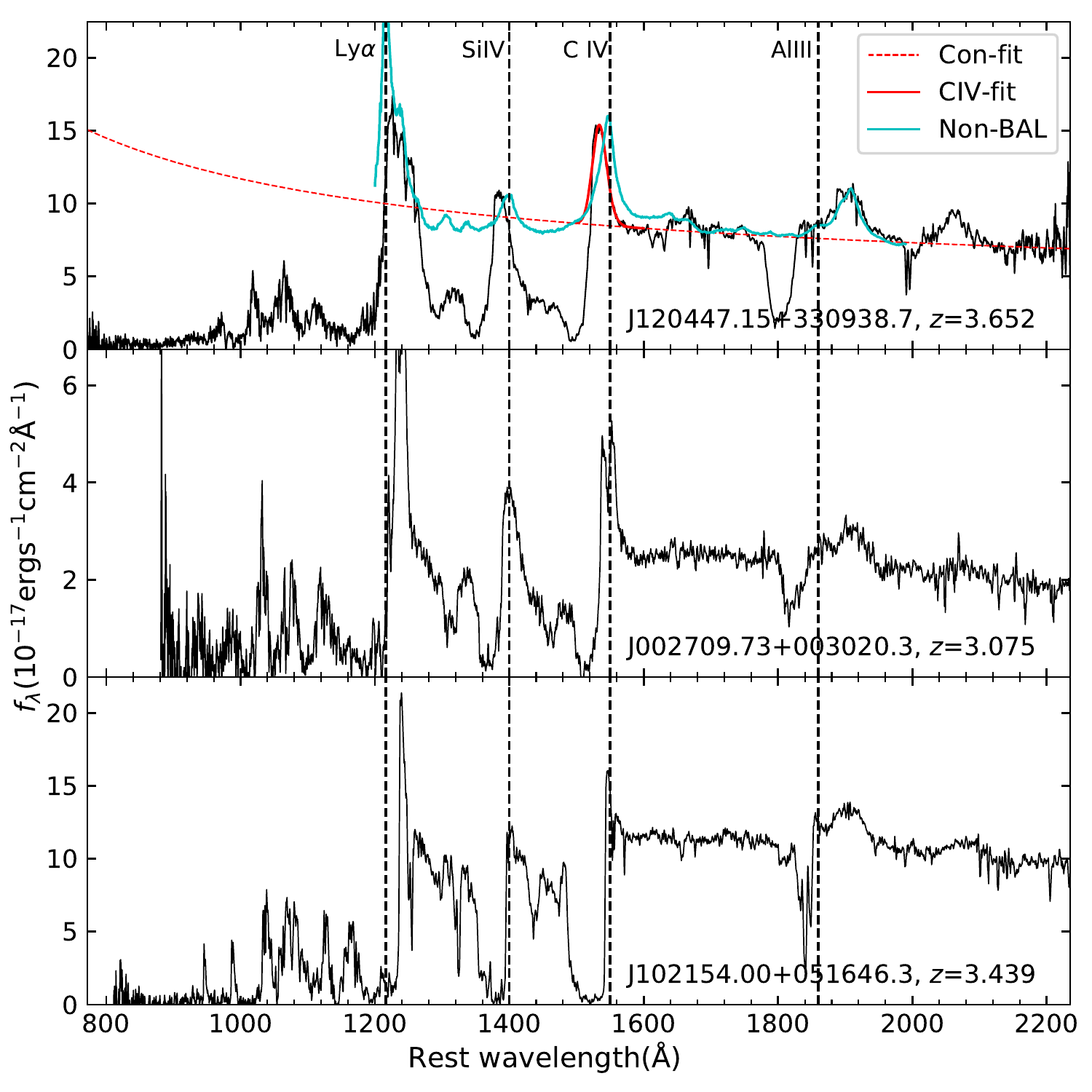}
       \caption{The top panel demonstrates continuum (red dashed) and non-BAL template (cyan) fits. The red solid line is a Gaussian fit of the \civ\ line. The middle/bottom panels show the matched spectra with similar redshift, AI(\civ), and spectral shape.}
\label{J120447matchspec}
\end{figure}

Similar to previous studies (e.g., \citealp{Grier16,Yi19a}), we adopt a reddened power-law model to fit the continuum  with a nonlinear least-squares fitting algorithm. We modeled the continuum in each spectrum using a SMC-like reddened power-law function from \citet{Pei92} with three free parameters including the amplitude, power-law index, and extinction coefficient. 
The initial continuum-fit windows are composed of relatively line-free (RLF) regions. However, these default windows do not always reasonably sample the continuum for all quasars in the sample, so they are adjusted where necessary to reach an acceptable fit. In particular, apparent emission/absorption lines present in the default windows among all spectra have been excluded. 

We masked pixels flagged by ``and-mask'' from SDSS pipeline before the continuum fit. All spectra are then converted to the rest frame using the redshifts determined by the \oiii, \hb, or \mgii\ line from the near-IR spectra. 
Some emission and/or absorption features appear to be frequently present in the same fitting regions. Therefore, we adopt a sigma-clipping algorithm that consists of fitting the reddened power-law function to the RLF windows for each spectrum, and then rejecting data points that deviate by more than 3$\sigma$ from the continuum fit for each pixel in each window.  The final continuum fit is obtained by re-fitting the remaining data points. We do not attach physical meaning to the values of $E(B - V )$ derived from our fits since there is a degeneracy between the UV spectral slope and intrinsic reddening. The uncertainty of the continuum fit is obtained via a Monte Carlo approach through randomizations of spectral errors in the RLF windows. 
Then, quantities such as AI(\civ), minimum/maximum trough velocities, and EW, are measured based on the reddened power-law fits (see Table \ref{balqso_dr12}). 

\subsection{ Flux calibration of near-IR spectra} \label{scaling_intro}

\begin{figure} 
  \center{}
      \includegraphics[width=8.5cm, height=7cm, angle=0]{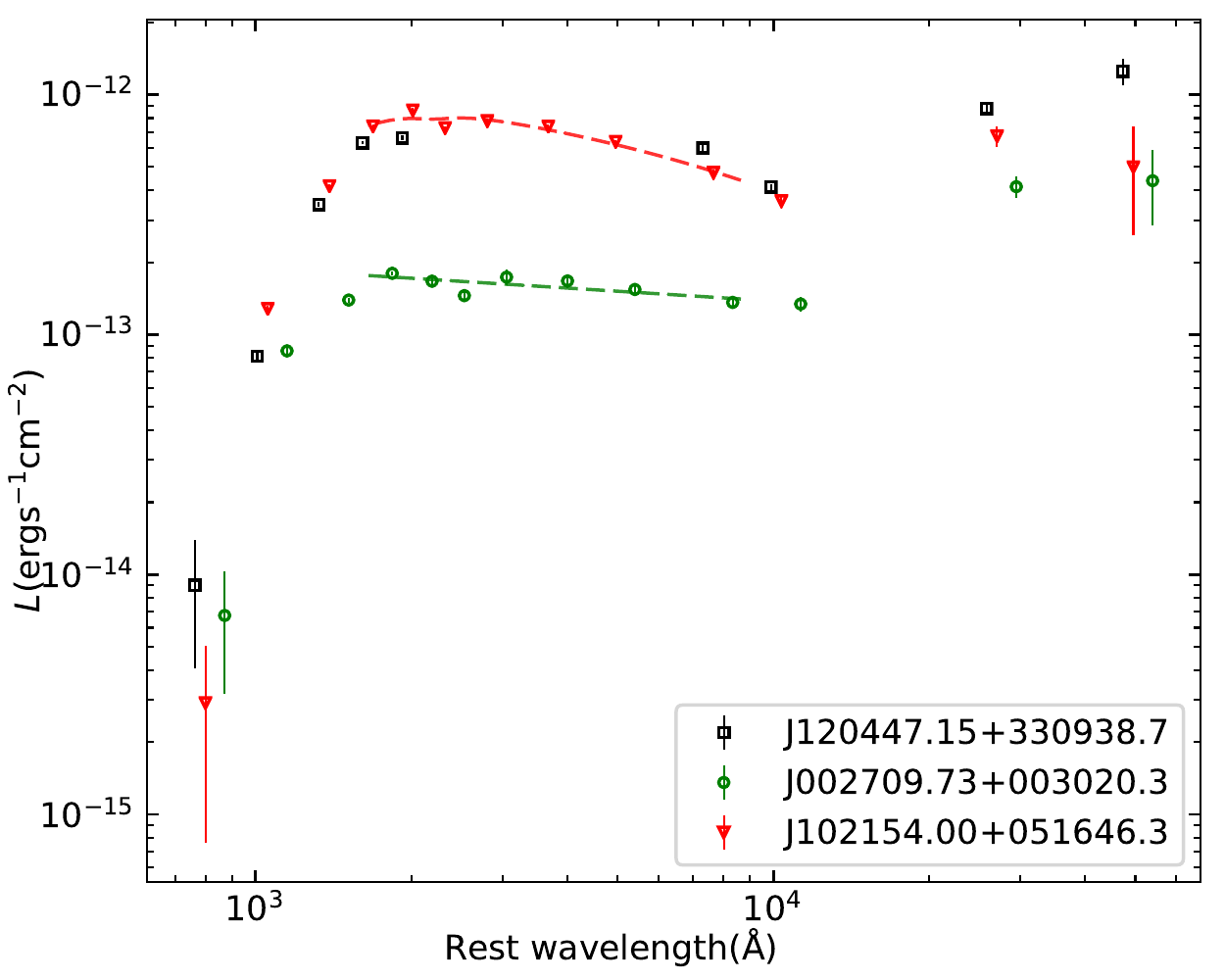}
       \caption{SEDs of the three quasars constructed from SDSS, UKIDSS  and WISE photometric magnitudes.  The dashed lines show reddened power-law (red) and power-law (green) fits for the two matched quasars between 1600\AA~ to 9000 \AA. }
   \label{J1204sed}
\end{figure}

The near-IR spectroscopic observations were carried out under non-photometric conditions. 
In addition, the cross-dispersion mode of spectroscopy may lead to a large uncertainty for near-IR spectral flux calibration, so an absolute flux calibration is required for reliable measurements of QSO properties. 
Since quasars usually show less variability at redder wavelengths, we use the converted photometric flux at SDSS-$i/z$, 2MASS-J/H/K if existed, and WISE-1/2 to scale the spectral flux. 
 Although the photometric and spectroscopic data were not simultaneously observed, for simplicity, we assume no significant continuum variability between the photometric and spectroscopic epochs for all QSOs in the sample.  For the BAL QSOs without J/H/K magnitudes in our sample, we select reference BAL QSOs with J/H/K magnitudes and similar optical spectral shapes to our targets from DR14. We then  use the models that are best fitted for the continua of the reference QSOs to fit our targets with SDSS-$i/z$ and WISE-1/2 magnitudes.  

As an example demonstrated in Figure \ref{J120447matchspec} and \ref{J1204sed}, two BAL QSOs were finally chosen as a reference to guide the local SED fit for the target (J120447.15+330938.7) without near-IR photometric observations. 
SEDs in the two matched BAL QSOs appear to be redder than non-BAL QSOs, most likely due to intrinsic reddening and Ly$\alpha$-forest absorption at high redshift. 
We only account for the local SED fit between 1600 and 10000 \AA\ in the rest frame, where a power-law or reddened power-law continuum shape is expected. A long wavelength cut is necessary since the hot-dust reprocessing starts to dominate the QSO SED at $\lambda >$ 10000 \AA. 
The local SEDs of the two reference QSOs can be well fitted by the power-law and reddened power-law models, respectively (see Figure~\ref{J1204sed}). 
Finally, we fit the local SED of our target using the above two models that are applied to the SED fitting process of the two reference QSOs in the same wavelength range (see Figure \ref{J120447hb_fit}). Their differences are taken as propagated uncertainties for the placement of the continuum. As the uncertainty is dominated by the above procedures,  we did not account for additional uncertainties, such as the uncertainty generated from  Monte Carlo simulations.

In general, we found that the 5100 \AA\ flux density derived from a power-law fit is quite close to that from a reddened power-law fit while a larger deviation may appear at 3000 \AA\ for these quasars without near-IR photometric observations in the sample.  
Thus, the continuum placement at 5100 \AA\ is our preferential choice for the conversion to the bolometric luminosity using the nominal  factors (5.18 for 3000 \AA\ and 9.26 for 5100 \AA, respectively; see \citealp{Shen:2011}). 
Finally, the absolute flux calibration is corrected by the Galactic extinction using  $R$v = 3.1 Milky Way extinction model (Cardelli et al. 1989) and a corresponding $A$v value \citep{Schlafly11}.

\subsection{The emission-line measurements}  \label{sec:fitting}

 \begin{figure*}
  \center{}
      \includegraphics[width=18cm, height=8cm, angle=0]{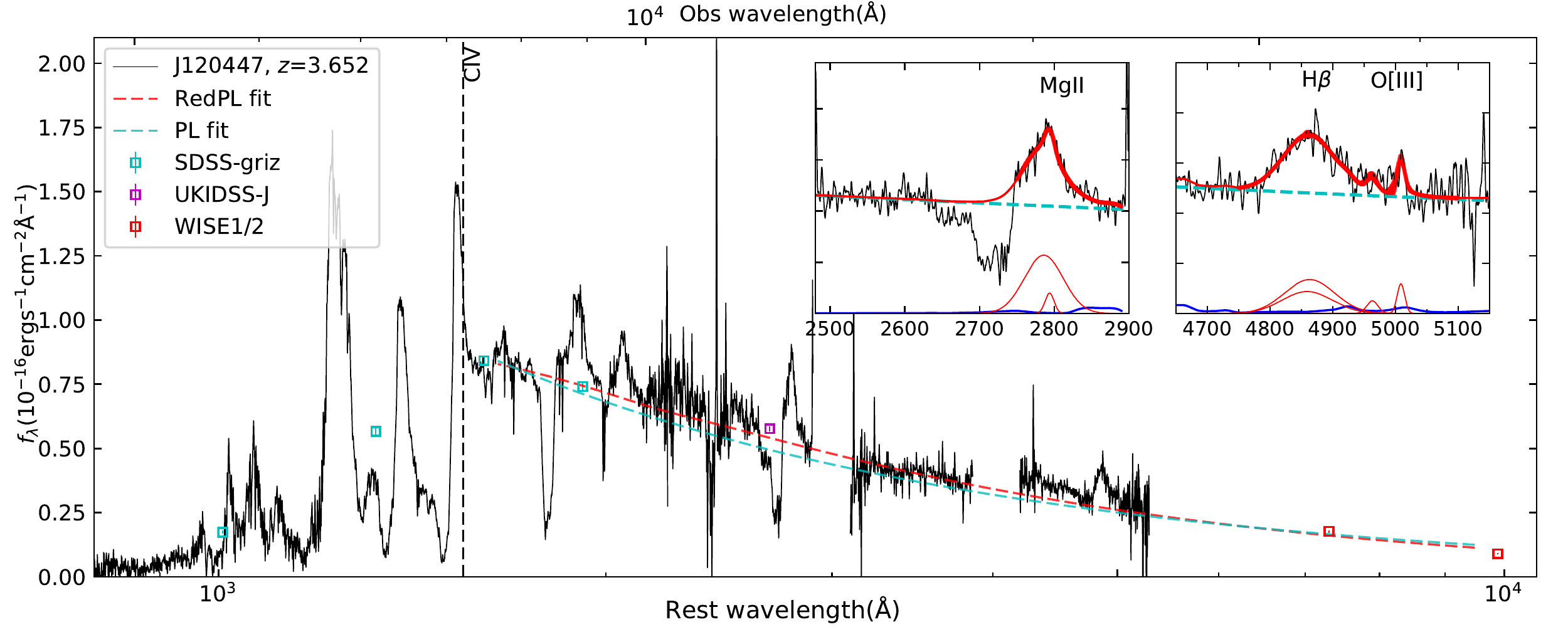} 
       \caption{An example of the SED fit using power-law (PL) and reddened PL  functions, respectively, based on photometric (cyan and red squares) data; black dashed line show an apparent blueshift compared to the systemic redshift determined by the \oiii5007 emission line. The cyan, magenta, and red squares represent photometric data from SDSS, UKIDSS, and WISE, respectively. Inset panels show spectral fits around the \mgii~ and H$\beta$ emission lines. The cyan dashed lines represent the power-law fit for the continuum, blue lines depict the fitted UV/optical \feii\ components, and the thin red lines represent all broad and narrow Gaussians for the fits around \mgii, \hb\ and \oiii\ emission lines. } 
 \label{J120447hb_fit}
\end{figure*}

\begin{table*}
\caption{Spectral measurements and derived black hole properties based on P200/TripleSpec observations}
\begin{tabular}{lcccccccccc}
\hline
Name &  FWHM$_{\rm \mgii}$ & $\log L_{\rm 3000\AA}$ & FWHM$_{\rm{H}\beta}$ & $\log L_{5100}$ & FWHM$_{\rm [O III]}$ & $\log L_{\rm{bol}}$ & $\log \mbh$ & $\log \lambda_{\rm Edd}$  \\
SDSS J & (km\ s$^{-1}$) & (erg s$^{-1}$) & (km\ s$^{-1}$) & (erg s$^{-1}$) & (km\ s$^{-1}$) & (erg s$^{-1}$) &($M_\odot$) &  \\
\hline

0140$-$0839 & 4735 $\pm$ 2430 &  47.25 $\pm$ 0.01 & 5348 $\pm$ 697 &  46.99 $\pm$ 0.01 &  1870 $\pm$ 500 & 47.96 $\pm$ 0.01  &  9.86 $\pm$ 0.11 &  $-$0.02 $\pm$   0.12 \\ 
0150+0041 & 5668 $\pm$ 105 &  46.95 $\pm$ 0.01& 5225 $\pm$ 780 &  46.74 $\pm$ 0.01 &  1410 $\pm$ 80 & 47.71 $\pm$ 0.01  &  9.38 $\pm$ 0.12 &  0.01 $\pm$   0.13 \\ 
0216$-$0921 & 2550 $\pm$ 650 &  46.99 $\pm$ 0.03 & 3729 $\pm$ 80 &  46.77 $\pm$ 0.03 &  - &  47.74 $\pm$ 0.01 &  9.38 $\pm$ 0.03 &  0.25$\pm$ 0.05 \\ 
0746+3014 & 4517 $\pm$ 2078 &  47.13 $\pm$ 0.09 & 7018 $\pm$ 900 &  46.90 $\pm$ 0.01 &  1300 $\pm$ 360 & 47.87 $\pm$ 0.01  &  10.05 $\pm$ 0.19 &  $-$0.29 $\pm$   0.19 \\ 
APM08279 & - &  - & 5020 $\pm$ 38 &  46.92 $\pm$ 0.5 &  - & 47.89 $\pm$ 0.5 &  9.80 $\pm$ 0.30 &  0.001 $\pm$ 0.30 \\ 
0837+4828 & 4244 $\pm$ 260 &  46.83 $\pm$ 0.08 & 5197 $\pm$ 1630 &  46.63 $\pm$ 0.03 &  1436 $\pm$ 240 & 47.59 $\pm$ 0.03  &  9.65 $\pm$ 0.25 &  $-$0.17 $\pm$   0.23 \\ 
0844+0503 & 3850 $\pm$ 830 &  47.44 $\pm$ 0.01 & 5812 $\pm$ 35 &  47.25 $\pm$ 0.01 &  2100 $\pm$ 90 & 48.21 $\pm$ 0.01  &  10.06 $\pm$ 0.01 &  0.03 $\pm$ 0.03 \\ 
0919+1938 & 3800 $\pm$ 1690 &  46.82 $\pm$ 0.1 & 5214 $\pm$ 380 &  46.70 $\pm$ 0.03 &  1400 $\pm$ 690 & 47.67 $\pm$ 0.03  &  9.70 $\pm$ 0.07 &  $-$0.14 $\pm$  0.11 \\  
1032+5140 & - &  - & 3193 $\pm$ 820 &  46.63 $\pm$ 0.02 &  - & 47.59 $\pm$ 0.02  &  9.23 $\pm$ 0.21 &  0.25 $\pm$  0.32 \\  
1150+2819 & 6024 $\pm$ 5450 &  47.24 $\pm$ 0.02 & 4071 $\pm$ 60 &  47.02 $\pm$ 0.02 &  1069 $\pm$ 35 & 47.99 $\pm$ 0.02  &  9.64 $\pm$ 0.02 &  0.24 $\pm$ 0.04 \\  
1204+3309 & 4928 $\pm$ 188 &  46.93 $\pm$ 0.04 & 5785 $\pm$ 165 &  46.88 $\pm$ 0.02 &  1400 $\pm$ 560 &  47.84 $\pm$ 0.02 &  9.87 $\pm$ 0.03 &  $-$0.14 $\pm$  0.04 \\ 
1210+1741 & - &  - & 6476 $\pm$ 960 &  47.07 $\pm$ 0.01 &  - &  48.04 $\pm$ 0.01 &  10.07 $\pm$0.04 &  $-$0.14 $\pm$ 0.05 \\ 
1237+0841 & 4840 $\pm$ 2030 &  47.10 $\pm$ 0.01 & - &  - &  - &  47.81 $\pm$ 0.01 &  9.77 $\pm$ 0.4 &  $-$0.07 $\pm$ 0.4  \\ 
1259+6101 & 2400 $\pm$ 4000 &  46.99 $\pm$ 0.01 & 5540 $\pm$ 2250 &  46.83 $\pm$ 0.01 &  1680 $\pm$ 310 &  47.80 $\pm$ 0.01 &  9.81 $\pm$ 0.22 &  $-$0.13 $\pm$  0.24 \\ 
1415+1129 & 4593 $\pm$ 900 &  46.71 $\pm$ 0.03 & 3915 $\pm$ 125 &  46.49 $\pm$ 0.03 &  1390 $\pm$ 86 &  47.46 $\pm$ 0.03 &  9.66 $\pm$ 0.05 &  0.002 $\pm$ 0.18 \\ 
1503+3641 & 3428 $\pm$ 820 &  47.20 $\pm$ 0.02 & 5336 $\pm$ 62 &  46.88 $\pm$ 0.02 &  1387 $\pm$ 86 &  47.85 $\pm$ 0.02 &  9.80 $\pm$ 0.02 &  $-$0.07 $\pm$ 0.08 \\ 
\noalign{\smallskip} \hline \noalign{\smallskip}
0122+1216 & 4210 $\pm$ 160 &  46.79 $\pm$ 0.04 & - &  - &  - &  47.50 $\pm$ 0.04 &  9.47 $\pm$ 0.06 &  $-$0.1  $\pm$ 0.10 \\ 
0928+5340 & 4408$\pm$ 190 &  46.99 $\pm$ 0.1 & - &  - &  - &  47.70 $\pm$ 0.1 &  9.64 $\pm$ 0.06 &  $-$0.05 $\pm$ 0.06 \\ 
1048+4407 & 4162 $\pm$ 357 &  46.97 $\pm$ 0.05 & - &  - &  - &  47.68 $\pm$ 0.05 &  9.58 $\pm$ 0.09 &  $-$0.01 $\pm$ 0.15 \\ 
1335+4101 & 3010 $\pm$ 2200 &  47.06 $\pm$ 0.01 & - &  - &  - &  47.78 $\pm$ 0.01 &  9.34 $\pm$ 0.48 &  0.315 $\pm$ 0.50 \\ 
1510+5148 & 4515 $\pm$ 2250 &  46.51 $\pm$ 0.05 & - &  - &  - &  47.24 $\pm$ 0.05 &  9.43 $\pm$ 0.36 &  $-$0.31 $\pm$ 0.41 \\ 
1638+1500 & 3670 $\pm$ 970 &  47.02 $\pm$ 0.1 & - &  - &  - &  47.74 $\pm$ 0.1 &  9.50 $\pm$ 0.25 &  0.12 $\pm$ 0.26 \\ 

\hline
\end{tabular}
\emph{Note.} $L_{\rm{bol}}$, $\mbh$ and $\log \lambda_{\rm Edd}$ are derived from the broad H$\beta$ emission line if available; otherwise, they are estimated using the \mgii\ emission line. The error bars of BH masses are the measurement uncertainties derived from 100 Monte Carlo simulations (not including systematic uncertainty of the scaling relation, which typically is about 0.3 dex). 

\label{tab:prop}
\end{table*}

Emission line properties are measured by fitting the local regions around the \hb\ and/or \mgii\ emission lines with a pseudo continuum+line model (see inset panels in Figure~\ref{J120447hb_fit}). Strong telluric-absorption windows are masked out in the near-IR spectra before fitting. 
For consistency, we adopt a similar method of continuum fitting and emission-line modeling from  \citet{Schulze17}, which is illustrated as follows: 

\begin{enumerate}
\item
The continua around \mgii\  and \hb\ emission lines are fitted by the model including a power-law (PL) function+optical/UV \feii\ templates \citep{Boroson:1992,Salviander07}. We do not consider the Balmer continuum in this work. 

\item
The broad emission-line fitting is then based on the pseudo-continuum subtracted spectrum. No more than three broad and one narrow  Gaussians are adopted to fit broad emission lines. We mask out absorption troughs during the fitting around the \mgii\ emission-line if its profile is affected by these troughs. 

\item
Due to the weakness of the \oiii\  emission line in the sample and spectral quality, the \oiii\ emission line is fitted by one Gaussian for most quasars in the sample.  The upper limits on \oiii\ blueshift and \oiii\ FWHM are set by 1000 and 2000 $\kms$, respectively, to account for possible broad \oiii\ components that may be associated with quasar outflows. 

\item
Since the widely used \citet{Boroson:1992} template may fail to provide a satisfactory \feii\ fit for all spectra, following \citet{Bischetti17}, we add additional two Gaussians during the fit for those objects with strong \feii\ emission at $\sim$4940 \AA\ and $\sim$5040 \AA, respectively.  

\end{enumerate}

\begin{figure*}
\center{}
      \includegraphics[width=16cm, height=10cm, angle=0]{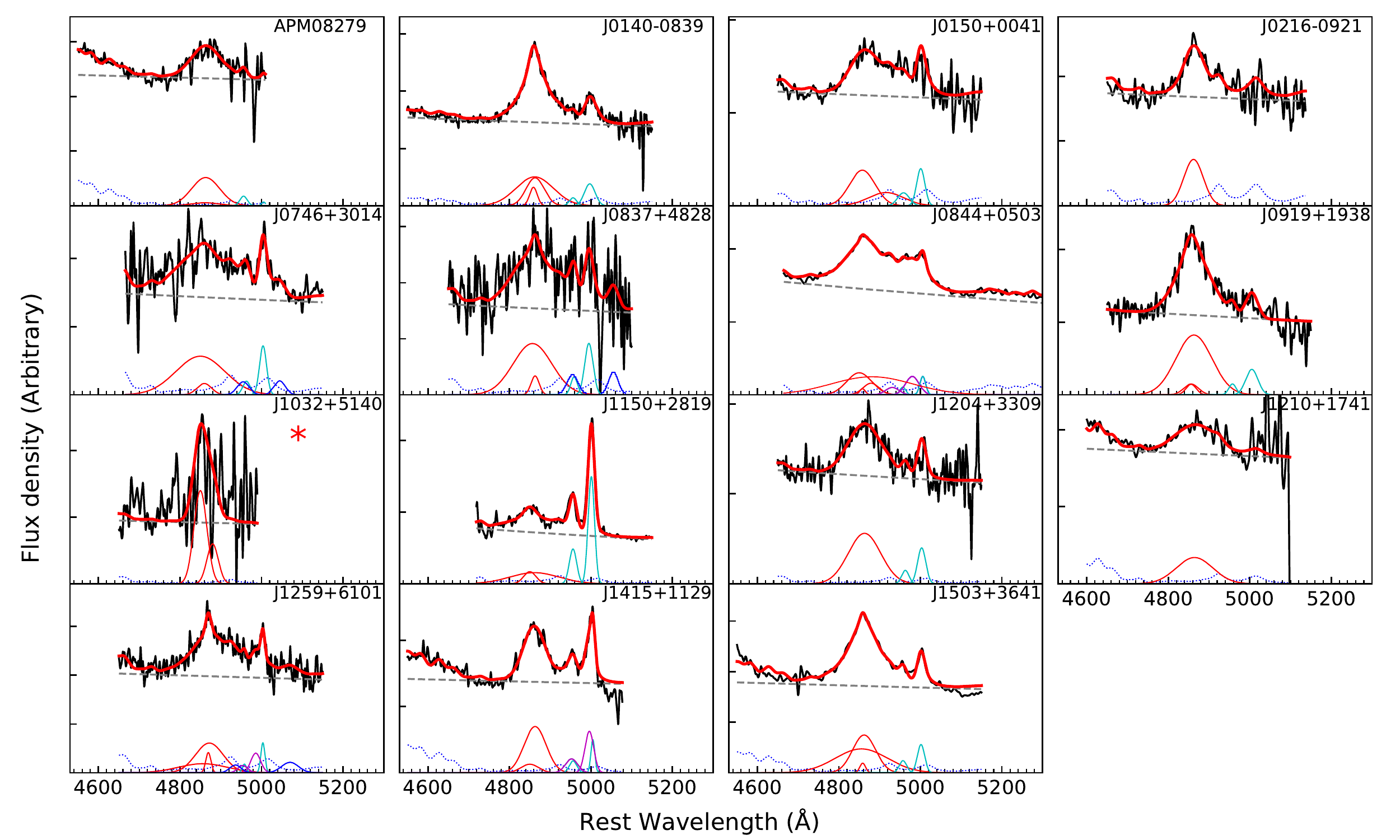}
      \includegraphics[width=16cm, height=12cm, angle=0]{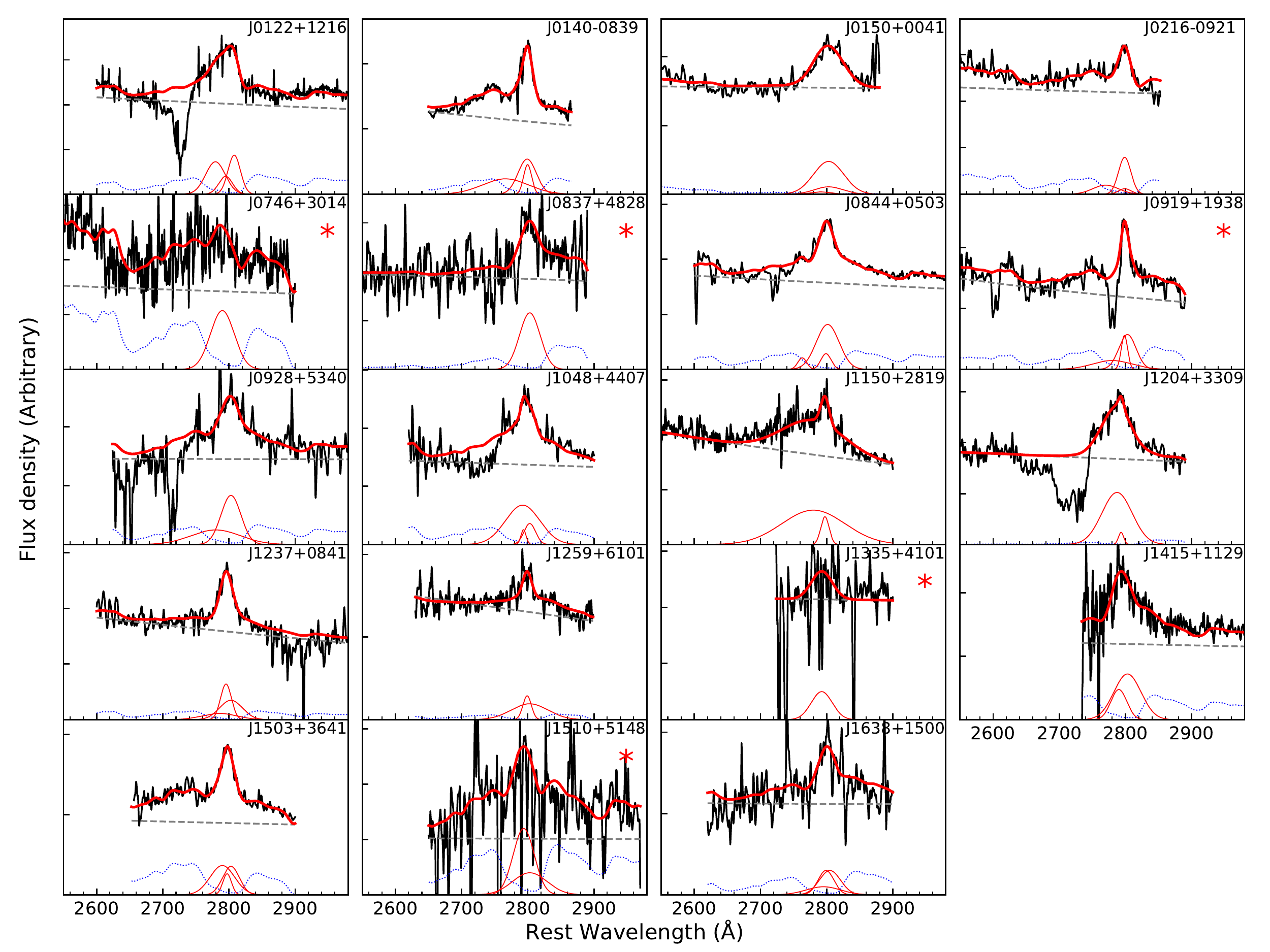}
       \caption{ Spectral fits around the \hb\ + \oiii\ (upper panel) and \mgii\ (lower panel) emission lines in our sample. The model of \hb\ emission line includes the power-law continuum (grey dashed), optical \feii\ template (dotted blue), multiple Gaussians for the broad \hb\ line (thin red), the \oiii$\lambda$4960/5008 doublet (cyan and magenta lines, in which magenta lines represent broad FWHM components of the \oiii$\lambda$4960/5008 doublet lines), and additional two broad Gaussians for \feii\ bumps at 4940 \AA\ and 5040 \AA\ (blue line) if exist. The model of \mgii\ emission line includes the power-law continuum (grey dashed), UV \feii\ template (dotted blue), multiple Gaussians for the broad \mgii\ line (thin red). The red stars refer to objects with low S/N spectra and/or absorption attached on the line.}
\label{hb_line_fits}
\end{figure*}

The upper limit on the velocity width (2000 km s$^{-1}$) used for the \oiii\ line fit is obviously larger than the typical velocity width ($<900$ km s$^{-1}$) of  narrow emission lines (NELs). However, this  upper limit is reasonable since  signatures of NEL outflows appear to be common in the luminous quasar population, particularly at high redshifts (e.g., \citealp{Netzer04,Harrison14,Bischetti17}). 
The wavelength separation between the \oiii$\lambda$5007 and \oiii$\lambda$4960 component has been constrained to the range from 45~\AA\ to 50~\AA\ in the rest-frame, with a ratio of their amplitudes constrained between 2.5:1 and 3:1. 
Limited by the quality of these near-IR spectra, we use only one Gaussian to fit the \oiii$\lambda$5007 emission line for the majority in the sample except for three QSOs with apparent broad \oiii\ profile  and high S/N (J0844+0503, J1259+6101, and J1415+1129, see Figure~\ref{hb_line_fits}).  
In particular, we applied two methods to decompose the plateau-like profile around the \hb\ + \oiii\ region in J0844+0503. One is followed by the above procedures; the other is based on the introduction of extremely blueshifted ($>$1700 \kms) and broad \oiii\ components (e.g., \citealp{Coatman2019,Perrotta2019}). The two methods yield a similar \oiii\ EW including both narrow and broad Gaussians, but we favor the latter since this quasar has sub-relativistic BAL outflows that may affect its NEL region (e.g., \citealp{Faucher:2012}). 
For the measurement of FWHM of the \hb\ emission line, we consider  all broad Gaussians (up to three components) during the fitting process. 

Spectral fits around the \mgii\ emission line, however, are more challenging for various reasons. First, there is no available \mgii-based size-luminosity relation such as that derived from the \hb\ line from reverberation mapping campaigns (e.g., \citealp{ Kaspi2000, Bentz2013}), indicating BH mass estimates using the \mgii\ line may be less reliable than those using the \hb\ line. Second, the \mgii\ line profile is potentially affected by absorption in our sample, especially for these objects showing deep and wide \civ-BAL troughs. Although the \mgii-BAL troughs are generally weaker and narrower than those found in \civ, they are often attached on the \mgii\ emission-line profile, making their identifications more difficult than those of \civ-BAL troughs. 
Third, \citet{Popovic19} reported that the \mgii\ broad emission line appears to originate from two subregions: one contributes to the line core that is probably virialized, and the other is associated with an emitting region characterized with outflows and inflows nearly orthogonal to the disc, which mostly contributes to the emission of the \mgii\ broad line wings. Fourth, whether or not considering the narrow component during the \mgii\ line decomposition may significantly affect the FWHM measurement of the broad \mgii\ line. Other challenges arise from different recipes of the single-epoch relation used in previous studies, the possibility of \mgii\ emission-line blueshift (cannot be determined without the aid of \hb), and the correction for the \mgii-based BH mass (e.g., \citealp{Woo2018}), and these issues need to be considered as well. Therefore, some caveats must be kept in mind when using the \mgii\ broad line for the BH mass estimation, especially in these quasars with extremely broad \mgii\ emission lines (FWHM $>$ 6000 $\kms$, e.g., \citealp{Yi19b}) and/or strong blue asymmetric profile in the \mgii\ emission line  (e.g., \citealp{Plotkin15}).

To mitigate the contamination of absorption and reconstruct the \mgii\ emission line in a reasonable sense, we mask out these absorption regions through visual inspection. The peak position of the \mgii\ emission line is constrained to the wavelength range between 2796~\AA\ and 2803~\AA. No more than three broad and one narrow Gaussian components are taken into account during the \mgii\ emission-line fit. Since the spectral quality does not allow us to decompose the \mgii\ line in a reliable way, we did not exclude the narrow component for the FWHM-\mgii\ measurement, even if it exists in some cases. Therefore, the FWHM of the \mgii\ emission line is determined by combining both the broad and narrow (if exists) Gaussian components after the spectral fitting. This would not bias our statistical results since it causes a  systematic underestimation of FWHM-\mgii\ for each quasar in the sample. 

The whole spectral fitting process for an individual object is demonstrated in Figure~\ref{J120447hb_fit}, in which we scaled the optical/near-IR spectra by a ratio between the spectral flux and the continuum fit based on the converted photometric flux. The local SED fits derived from both the power-law and reddened power-law functions are demonstrated by the cyan dashed and red dashed lines, respectively. All the spectral fitting results around the \mgii\ and/or \hb\ emission lines in the sample are shown in Figure~\ref{hb_line_fits}. 

\subsection{BH mass estimate } \label{BH_mass_estimate}
At high redshift, the single-epoch scaling relation is the primary method adopted to estimate BH mass in quasars. A general equation of the relation can be expressed as follows: 
\begin{equation}
 log ( \frac{M_{\rm BH}} {M_\odot}  ) = a +  b\times log(\frac{\lambda L_{\lambda}} {10^{44}   ergs^{-1}}) + 2log (\frac{\rm FWHM} { kms^{-1}}) 
\label{eq1}
\end{equation}

There are several parameter configurations ($a,b$) in terms of the scaling relation from the literature (for details, see Section 3.7 in \citealp{Shen:2011}). As a sample study, it is mandatory to adopt the same method to avoid introducing additional  uncertainties for the measurements of each object. 
Throughout this work, for \hb, we use the configuration of $a=0.91,b=0.5$ following \citep{Vestergaard:2006}; for \mgii, we use the configuration of $a=0.86,b=0.5$ following \citep{Vestergaard2009}. 

To quantify the measurement errors, we randomize spectral errors in the line fitting window via 100 Monte Carlo simulations and consider the standard deviation as the measurement error for each quasar. 
The typical measurement errors are small, only $\sim$ 0.1 dex (see Table~\ref{tab:prop}). However, we do note that the main uncertainties are dominated by systematic errors, which amount to $\sim$ 0.4 dex or even higher according to previous  studies (e.g., \citealp{Shen:2011}).

\subsection{The composite spectrum of strong BAL QSOs } \label{composite_spec}
There are 7 QSOs ($\sim$32\%) in our sample having strong \civ-BAL features (\civ-BAL EW $>$ 60~\AA) with an average S/N $>$ 7 at 3000~\AA\ or 5100~\AA. All of them are LoBALs and exhibit large \civ-BEL blueshifts measured by the apparent \civ-BEL peak. This method yields only a lower limit on the \civ-BEL blueshift due to the blueshifted BAL effect. In addition, all the seven QSOs have  trough-velocity widths of $\Delta v>10000$ km s$^{-1}$ and BAL velocities of $v_{max}>20000$ km s$^{-1}$ along our line of sight (LOS), indicating powerful nuclear outflows that may be capable of affecting their host galaxies via an energy-conserving expansion process (e.g, \citealp{Faucher:2012}). 
We thus construct a composite spectrum from these QSOs to investigate similarities and differences in the continuum and emission-line properties compared with the non-BAL composite from \citet{Zuo:2015} matched in luminosity  and the non-BAL composite from \citet{VandenBerk01}. 

\begin{figure} 
\center{}
     \includegraphics[width=8.6cm, height=4.3cm, angle=0]{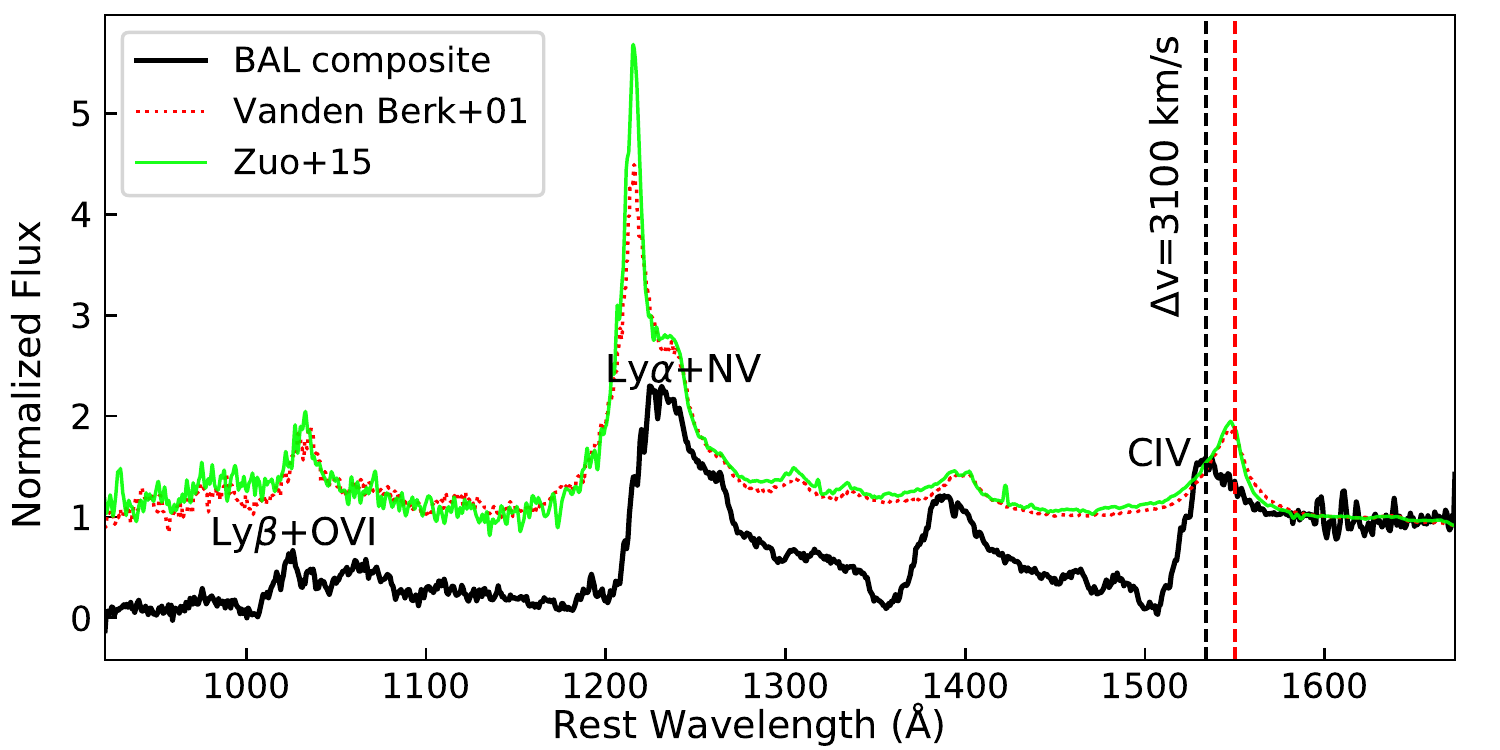}
      \includegraphics[width=8.6cm, height=4.3cm, angle=0]{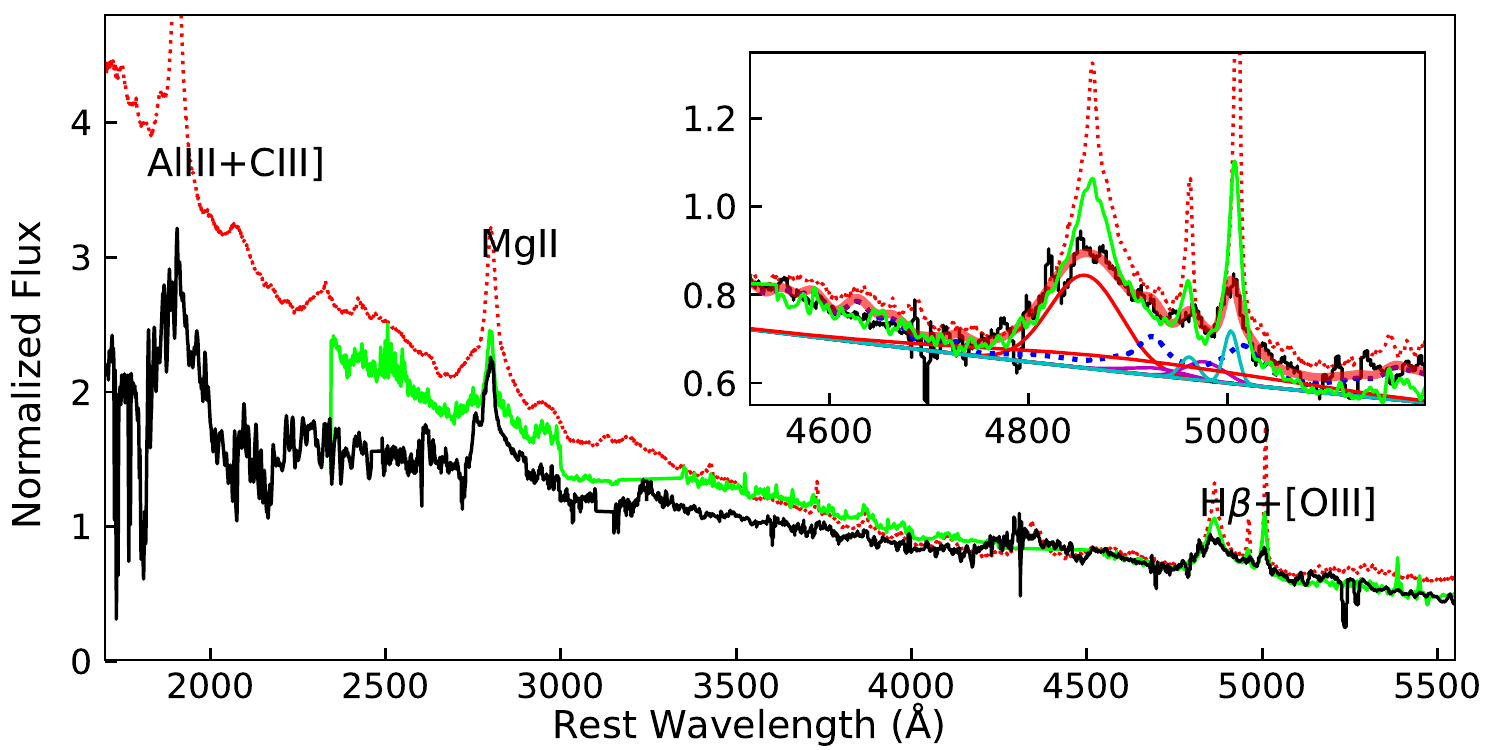}
       \caption{ The black lines show the composite spectra (smoothed by a 20-pixel Boxcar filter for clarity)  generated from seven LoBAL QSOs from our sample at $3<z<5$. The green and red spectra depict the non-BAL composites from \citet{Zuo:2015} and \citet{VandenBerk01}, respectively. The vertical dashed black line shows the apparent \civ-BEL blueshift with respect to \oiii. Clearly, our composite has a larger \civ-BEL blueshift and weaker \oiii\ emission than that from \citet{Zuo:2015} matched in luminosity. The inset panel shows the spectral fit around \hb, in which different-color components have the same meaning described in the caption of Figure \ref{hb_line_fits}.  }
\label{nir_composite}
\end{figure}

These optical/near-IR spectra used to construct the composite spectrum cover a wide wavelength range from 900~\AA\ to 5500~\AA\ in the quasar rest frame. 
All the selected spectra are normalized at 1600~\AA\ for the optical spectra and at 3600~\AA\ for the near-IR spectra. We use the geometric mean among these normalized spectra to produce stacked spectra as shown in Figure~\ref{nir_composite}, where the composite spectra at $900<\lambda <1700$~\AA\ (top panel) and $1800<\lambda <5500$~\AA\ (bottom panel) are generated by the optical and near-IR spectra, respectively. Since the near-IR spectra of the seven BAL QSOs have different redshifts ranging from 3.179 to 4.82, the composite spectra have some breaks due to the cut of telluric-absorption windows.

The BAL composite spectrum shows significant intrinsic reddening at $\lambda<2000$~\AA, though it appears to be largely free of reddening at $\lambda>4000$~\AA. The conspicuous differences  between the BAL and non-BAL composites at $\lambda_{\rm rest}<$1600 \AA\ could be due primarily to strong BAL absorption in \civ, \siiv, \nv+Ly$\alpha$, and \ovi+Ly$\beta$ at a similar velocity, plus stronger intergalactic medium  (IGM) absorption at higher redshift. After scaling the two non-BAL composites to the BAL composite, we found that all of them have a similar \feii\ emission-line profile at  4500$<\lambda_{\rm rest}<$4700~ \AA, suggesting that optical \feii\ emission may not depend on redshift, luminosity, and the presence of outflows. 
After performing a spectral fit to the \hb+\oiii\ region (see Figure~\ref{nir_composite}), we found a narrow, unshifted (cyan) and a broad, blueshifted \oiii\ doublets (magenta). 
In addition, the entire \hb\ emission-line profile can be well fitted by a single, almost unshifted Gaussian and a very broad, redshifted Gaussian, supporting  that the \hb\ emission line in BAL QSOs can serve as an equally reliable BH estimator as in non-BAL QSOs. The BAL composite shows an apparently wider \hb\ emission-line profile than the two non-BAL composites, probably signaling an intrinsic difference between the two quasar populations  (see the separation between Population A and B using FWHM-\hb\ from \citealt{Sulentic2000}). However, when taking a close-up look at the inset panel of Figure~\ref{nir_composite},  the broadening effect of the \hb\ emission line in the BAL composite appears  to be caused by the lack of peaky profiles in both \hb\ and \oiii\ compared to the two non-BAL composites. This implies that the presence of outflow may predominantly affect the narrow emission lines originated from large-scale regions. We discuss this issue in Section~\ref{discussion_section}. 
 

In good agreement with the finding from~\citet{Yi17}, our composite spectrum of strong BAL QSOs reveals a striking blueshift among all the high-ionization BELs ($\sim$3100 $\kms$, possibly higher due to the blueshifted BAL effect) and nearly black absorption troughs, which in turn provides evidence in support of outflows affecting the high-ionization BEL regions. Moreover, the composite spectrum shows significant reddening at $\lambda<3000$~\AA\ and weak \oiii\ emission. As a comparison, less than 25\% non-BAL QSOs from \citet{Zuo:2015} matched in redshift and luminosity show comparable weakness of the \oiii\ emission. In addition, the average \civ-BEL blueshift is  $\sim$600 \kms\ from \citet{Zuo:2015}, within which only one quasar has \civ-BEL blueshift larger than 2000 \kms\ (in private communication). It is worth noting at this point that all the above studies do not have simultaneous optical/near-IR spectroscopic data, which may potentially lead to larger uncertainties in the \civ-BEL blueshift than the values given above. However, the \civ-BEL blueshift could remain unchanged even if the \civ-BEL flux  varies dramatically from epoch to epoch (e.g., \citealp{Ross2019}).

\section{Comparison with other samples} \label{results_section}
In this section, we investigate the distributions of physical properties and compare them with other non-BAL and BAL samples matched within a similar luminosity and/or redshift range to our sample. 

\subsection{Comparison of direct measurements between BAL and non-BAL QSOs} \label{non_bal_sample}
  \begin{figure}
\center{}
      \includegraphics[width=8cm, height=14cm, angle=0]{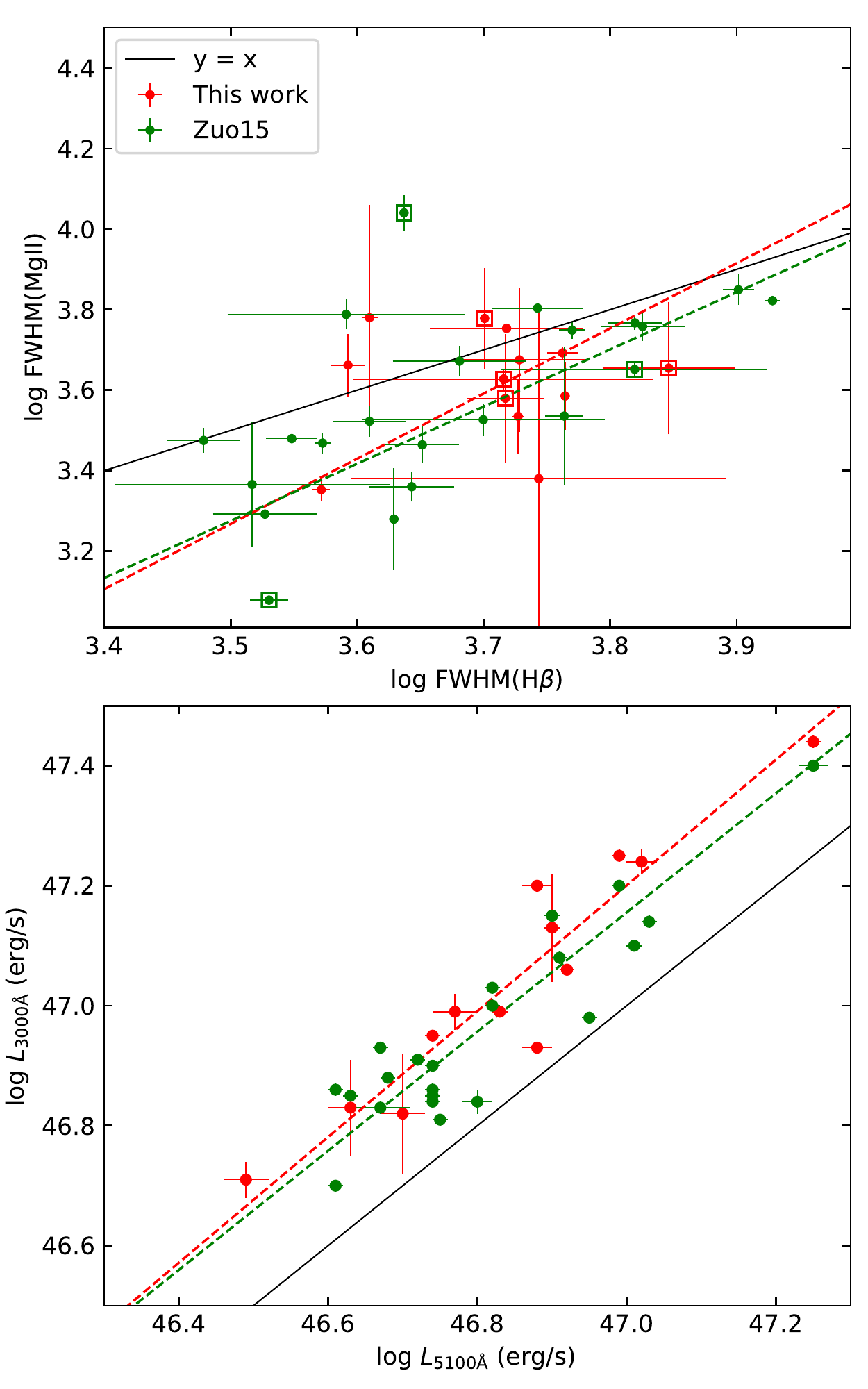}
       \caption{Top: FWHMs of \mgii\ versus \hb\ emission lines. Black line is the 1:1 distribution. The open red/green squares represent these objects with relatively low-quality fitting results  from the two samples. Bottom: Distribution of 5100~\AA\ monochromatic luminosity versus 3000~\AA\ monochromatic luminosity. Red and green dashed lines are the linear fits based on the orthogonal distance regression. }
\label{com_fwhm_hb_mg}
\end{figure}
Previous studies based on non-BAL quasar samples found that the FWHM distribution of the \mgii\ BEL is (linearly) correlated with that of the \hb\ BEL at a high significance level (e.g., \citealp{Trakhtenbrot:2012, Zuo:2015}), thus the \mgii\ BEL is usually considered as an alternative BH-mass estimator in good agreement with the \hb\ line in the non-BAL population.

There are 13 objects in our sample with simultaneous observations of the \mgii\ and \hb\ emission lines, which provide an opportunity to test whether this correlation holds in the luminous BAL QSOs at high redshift.  
Four quasars (marked by open red squares in the upper panel of Figure \ref{com_fwhm_hb_mg}) in our sample have relatively low-S/N spectra and/or broad absorption attached on the \mgii\ line, which may bias the statistical results. 
We checked the correlation after excluding them, and found that the results are consistent with the subsample of 13 objects via the Spearman test. Therefore, we use the entire subsample of 13 BAL QSOs in our analyses below.

We choose three non-BAL  comparison samples (\citealt{Zuo:2015}, \citealt{Coatman2017}, and \citealt{Vietri18}; hereafter Zuo15, C17, and V18, respectively) within a similar range of luminosity and redshift to our sample. The  sample of Zuo15 includes 24 non-BAL QSOs at a similar redshift and luminosity to our sample, among which 22 of them have the coverage of both the \hb\ and \mgii\ emission lines. The near-IR spectra from Zuo15 and our sample were obtained by the same instrument (P200/TripleSpec), largely alleviating uncertainties caused by different instruments for the comparisons. The comparison sample of C17 consists of 230 non-BAL QSOs at $1.5<z<4$, from which we select 19 within the same redshift and luminosity ranges to our sample ($z>2.56$ and $46.5<$ log$L_{5100}<47.3$). Additionally, we select 13 non-BAL QSOs at $z\sim3.3$ and log$L_{5100}\sim47$ from \citet{Vietri18} to construct the V18 comparison sample.

According to the single-epoch scaling relation, the BH mass measurement is proportional to the square of FWHM and root square of monochromatic luminosity (the former is an indirect measurement depending on sophisticated analyses as roughly mentioned in Section \ref{sec:fitting}), it is essential to investigate the distributions of the two quantities in our sample and compare them with other studies. As shown in the top panel of Figure \ref{com_fwhm_hb_mg}, the distribution of FWHM-\mgii\ versus FWHM-\hb\ in this work is broadly consistent with that from \citet{Zuo:2015} via orthogonal distance regression fits \footnote{https://docs.scipy.org/doc/scipy/reference/odr.html} (red and green dashed lines are the two fits for the two samples, respectively). Although there is no significant difference in the distribution of FWHM-\mgii\ from the two samples, as determined via a two-sample K-S test ($p_{\rm KS}=0.23$), but a difference may exist in the distribution of FWHM-\hb\ ($p_{\rm KS}=0.1$). 
In addition, both FWHM distributions of the two samples deviate from the one-to-one line (black), with more objects showing FWHM-\hb\ larger than FWHM-\mgii. However, the FWHM distribution between the two different broad emission lines from our sample is less correlated compared with Zuo15, which is confirmed via the Spearman test ($r=-0.09,p=0.775$ and $r=0.67,p=0.0006$ for our sample and Zuo15, respectively). The lack of correlations in our sample is likely due to strong nuclear outflows traced by BALs and/or BEL blueshifts, when noticing that both samples have a similar luminosity and redshift range.

Conversely, the two samples follow a similar linear relation in the distribution of 5100 \AA\ versus 3000 \AA\ monochromatic luminosity  (see red/green dashed lines in the bottom of Figure \ref{com_fwhm_hb_mg}). Specifically, we found strong correlations for both samples via the Spearman tests ($r=0.92,p=10^{-5}$ and $r=0.75,p=5\times 10^{-5}$ for this work and Zuo15, respectively), 
which are consistent with previous studies based on non-BAL QSOs. The similarity in the luminosity relation and difference in the FWHM relation between the BAL and non-BAL samples, to some extent, support the scenario where strong BAL outflows are affecting the (\mgii) BEL regions.

\subsection{Comparison of derived measurements with non-BAL QSOs matched in luminosity and redshift}
\label{comparison_match_z_L}

\begin{figure}
\center{}
      \includegraphics[width=8.5cm, height=7cm, angle=0]{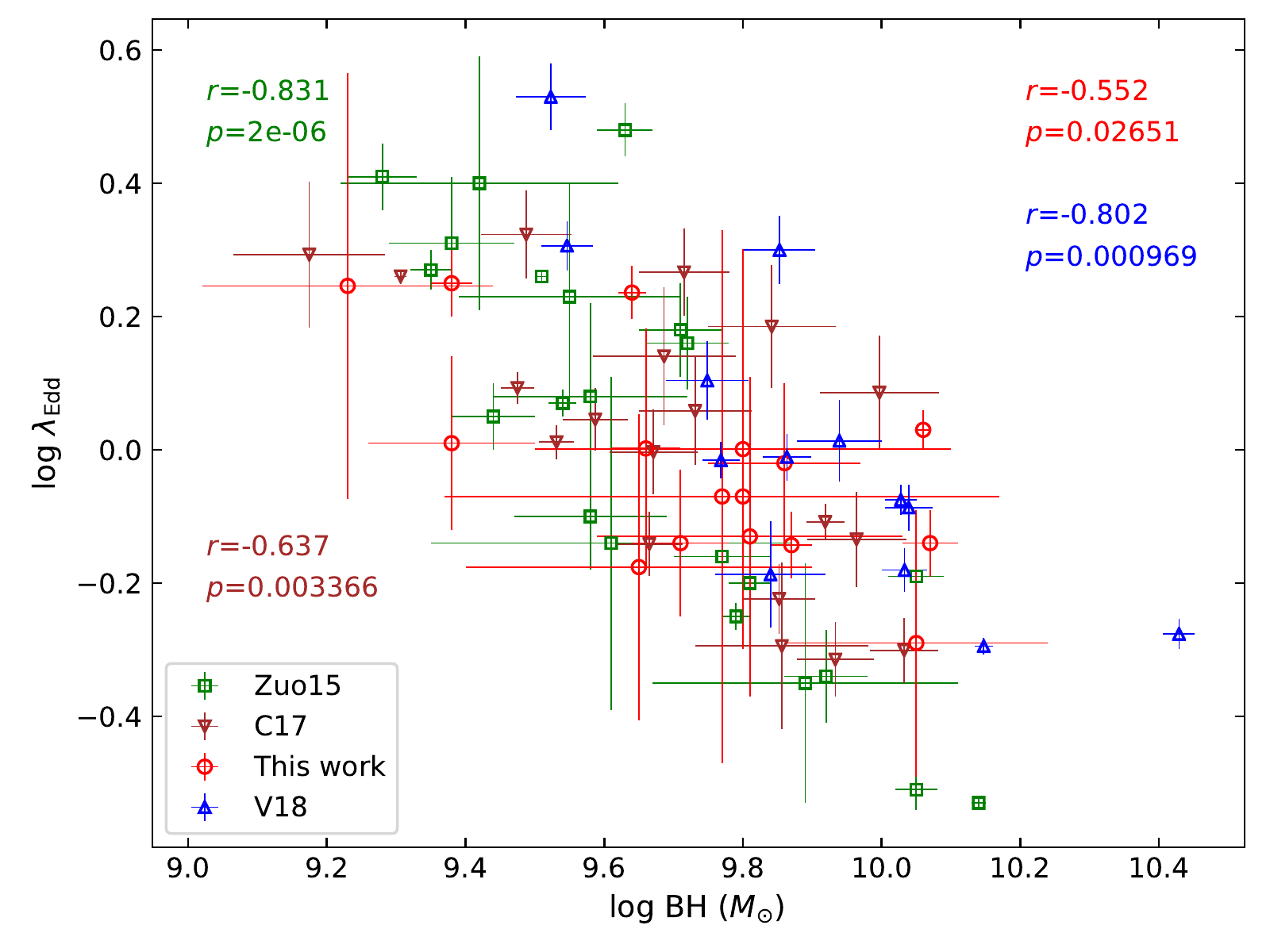}
       \caption{Distributions of Eddington ratio versus BH mass, in which the subset of the BAL sample (red) and three non-BAL samples have a similar redshift/luminosity range. All of the three non-BAL samples show strong correlations compared to a tentative correlation in the BAL sample. }
\label{com_bh_edd_z15}
\end{figure}

\begin{table*}
\center{}
\caption{Near-IR spectroscopic studies of BALQSOs at $z\gtrsim3$}
\begin{tabular}{lccccc}
\hline
& $L_{5100}^a$ & $L_{5100}^b$ & FWHM-\hb$^a$ & FWHM-\hb$^b$ & $\lambda_{\rm Edd}$   \\
\hline
$L_{3000}$  &($0.92,10^{-5}$) & ($0.75,5\times 10^{-5}$) & - & - &  -  \\
FWHM-\mgii  & - & - & ($-0.09,0.775$) & ($0.67,0.0006$) & -  \\
$M_{\rm BH}^a$  & - & - & - & -&   (0.552, 0.026) \\
$M_{\rm BH}^b$  & - & - & - & -  &  ($0.831,2\times10^{-6}$) \\
$M_{\rm BH}^c$  & - & - & - & -   & ($-0.637,0.003$)\\
$M_{\rm BH}^d$  & - & - & - & -  & ($-0.802, 9\times10^{-4}$) \\
\hline
\end{tabular}
\tablecomments{   Spearman test results ($r,p$) of direct ($L_{5100}$ and FWHM) and derived ($M_{\rm BH}$ and $\lambda_{\rm Edd}$) measurements from different samples, in which $a,b,c,d$ represent this work, Zuo15, C17, and V18, respectively. }
\label{tab_3disap_trans2}
\end{table*}

 Unlike the investigation and comparison of direct measurements in the above subsection, here we have to use derived measurements to explore the effect of outflows, due to the lack of simultaneous observations or the lack of reported \mgii\ and \hb\ lines  from C17 and V18 studies. We use the 16 BAL QSOs at $z<4$ from our sample and compare them with the above three non-BAL QSO samples based on near-IR spectroscopy in a similar luminosity and redshift  range to our sample (see Figure \ref{com_bh_edd_z15}). 
Based on the two-sample K-S test, there is no statistically significant difference in the distributions of Eddington ratio between the BAL and non-BAL samples; but potential differences may exist in the distribution of BH masses ($p_{\rm KS}=0.14$) between this work and Zuo15, tentatively hinting that the growth of SMBHs in our sample is regulated by BAL outflows. 
Since the K-S test is incapable of finding differences in the correlation strength between different parameters, and our sample, at first glance, appears to have no correlation in the distribution compared to apparent correlations in the other three matched comparison samples, we perform the Spearman test to quantify this difference. We found strong correlations in all the three non-BAL comparison samples at a highly significant level while only a tentative correlation in our sample (see  Figure~\ref{com_bh_edd_z15} and Table~ \ref{tab_3disap_trans2}).

The distributions of BH mass versus Eddington ratio above are unlikely biased because all of them are derived from the same BH mass estimator (\hb), which is largely free of the effects of intrinsic reddening and absorption. 
Despite a small number of quasars in the four samples, the difference in the correlation strength between the non-BAL and BAL QSO samples is unlikely caused by the difference in sample size, as the V18 (blue) sample consisting of fewer QSOs shows an even stronger correlation than those from C17 and our sample (the number of quasars with \hb-based BH masses in  Z15, C17, V18, and this work are 22, 19, 13, and 16, respectively). Therefore, we attribute such a difference to the effects of substantial outflows traced by strong BALs and large CIV-BEL blueshifts ubiquitously seen in our sample.

\subsection{Comparison of Eddington ratio with non-BAL QSOs at lower luminosities} \label{comparison_z}

There are another two non-BAL QSO samples with available estimated BH masses and Eddington ratios at $z\sim 4.8$, $z\sim2.4$ and 3.3 (\citealp{Trakhtenbrot11,Netzer07}). Although the two samples have a lower luminosity range, they can provide additional diagnostics for systematic comparisons between the BAL and non-BAL populations, particularly in the context of normalized accretion rate, namely Eddington ratio ($\lambda_{\rm Edd}$). 

\begin{figure} 
\center{}
      \includegraphics[width=8.5cm, height=7cm, angle=0]{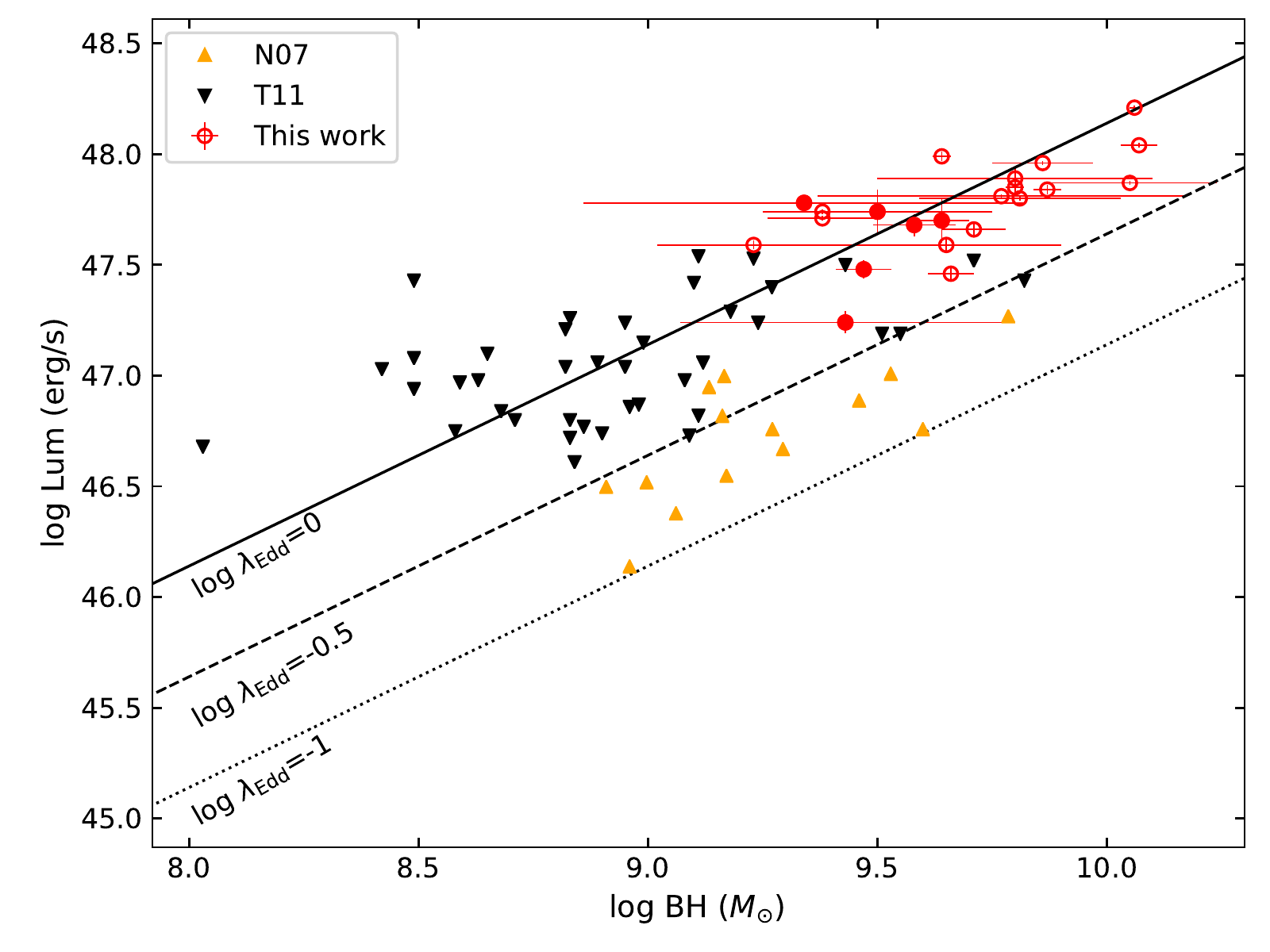} 
       \caption{Distributions of bolometric luminosity versus BH mass for the three samples, where dotted, dashed, and solid black lines represent log$\lambda_{\rm Edd}= -1,-0.5,0$, respectively. Filled/empty red circles are the BAL QSOs at $z>4.3$ and $z<4$ in our sample, respectively. It is clear that BAL QSOs on average do not have a higher Eddington ratio compared to the lower-luminosity non-BAL sample of T11 at $z\sim$4.8. }
\label{com_bh_lum_T11_N07}
\end{figure}

In this subsection, we use our entire sample consisting of 22 BAL quasars at $z>2.5$. 
BH masses from all the three samples were estimated using the \hb\ and \mgii\ emission lines at $z<4$ and $z>4$, respectively.  
Because the bolometric correction factor adopted in this work is about 1.5 times higher than that from the two non-BAL samples (see Section 5.1 from \citealp{Trakhtenbrot11}), we correct the distribution of bolometric luminosities by applying the same factor to the two comparison samples.  
As shown in Figure \ref{com_bh_lum_T11_N07}, there is a distinct separation between different quasar samples, with a  larger average BH mass in this work (this is quantitatively confirmed by a two-sample K-S test with $p_{\rm KS}=2\times 10^{-8}$). The vast majority of BAL QSOs in our sample are bounded by log$\lambda_{\rm Edd}=-0.5$ and log$\lambda_{\rm Edd}=0.5$, with a median Eddington ratio of log$\lambda_{\rm Edd}=-0.04$, a value that is approximately equal to that from T11.   However, our sample contains BAL quasars with significantly  higher luminosities and BH masses than T11 (see Figure \ref{com_bh_lum_T11_N07}). 

\begin{figure} 
\center{}
      \includegraphics[width=8.5cm, height=5cm, angle=0]{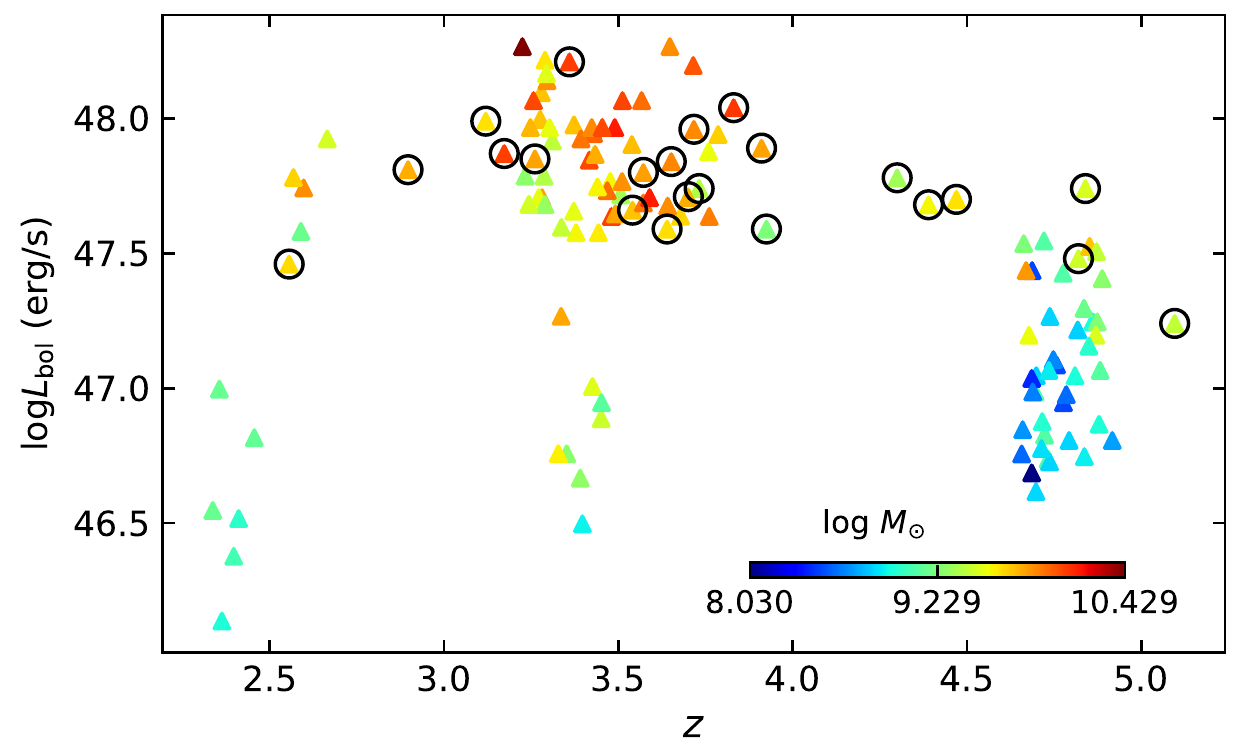} 
       \caption{Distributions of redshift versus bolometric luminosity from the BAL (triangles with overplotted circles) sample and non-BAL comparison samples constructed from Zuo15, C17, V18, N07, and T11, color coded by BH mass. }
\label{rds_lum_bh_bal_nonbal_all}
\end{figure}

There are six LoBAL QSOs at $z>4.3$ in our sample (filled red circles in Figure \ref{com_bh_lum_T11_N07}) with their BH masses estimated by the \mgii\ emission line. We compare the distributions of Eddington ratio between these six LoBAL QSOs and those from T11 sample, as they are also measured using the \mgii\ BH mass estimator. Based on the two-sample K-S test, we found that the two samples are likely drawn from the same distribution of Eddington ratio ($p_{\rm KS}=0.9$). Therefore, 
our observational results together indicate that BAL QSOs on average do not have higher Eddington ratios than those of non-BAL QSOs at similar luminosity and/or redshift, supporting the argument that high Eddington ratio is not sufficient for producing strong outflows (\citealp{Baskin2005}).

\subsection{An overview of the BAL and non-BAL samples} \label{overview_samples}
Figure \ref{rds_lum_bh_bal_nonbal_all} presents the distribution of redshift versus bolometric luminosity from our BAL sample and the aforementioned non-BAL comparison samples. The non-BAL QSOs with bolometric luminosity less than 47.6 erg~s$^{-1}$ in the logarithmic scale are from N07 and T11 as discussed in the above section. Note that at $z<4$ all the bolometric luminosities and BH masses are derived by the 5100 \AA\ continuum flux and the scaling relation for \hb, respectively, which in general, could provide more reliable measurements than those at $z>4$. At fixed luminosity, we do not see any clear trends with respect to the growth of SMBHs  either from our BAL sample or the augmented non-BAL sample from Zuo15, C17, V18, N07, and T11. However, limited by the sample size, we cannot draw any conclusions regarding the BH growth and redshift evolution. 

\section{Discussion} \label{discussion_section}
We present the observational results of our sample and systematic comparisons with other non-BAL samples matched in redshift and/or luminosity throughout section \ref{results_section}. Here, we systematically assess the role of nuclear outflows and their effects in conjunction with the reported findings from other BAL/non-BAL studies that are relevant to this work. We discuss the underlying link between BAL and BEL outflows, and propose an inhomogeneous outflow system filled with different-form QSO winds along different sightlines, which can mostly explain these observational results found from the BAL and non-BAL samples.

The lack of correlation between \mgii\ and \hb\ FWHMs, the marginal correlation between BH mass and Eddington ratio, the striking \civ-BEL blueshift and nearly black \civ-BAL trough seen in the composite, and most importantly the presence of BALs in our sample, together indicate that nuclear outflows are capable of affecting their BEL regions, particularly the high-ionization BEL region. This is also supported by recent studies (e.g., \citealp{Coatman2017,Vietri18}), where they found a strong correlation between \civ-BEL blueshift and \civ\ FWHM but no correlation between \civ-BEL blueshift and \hb\ FWHM. Additional  evidence can be found from some weak emission-line quasars, among which the BEL blueshift increases dramatically from \mgii\ to \civ\ (e.g., \citealp{Plotkin15,Yi19b}), supporting that nuclear outflows may exert a significant effect only on the high-ionization BELs. Combining other observational results from large sample studies, i.e. BAL QSOs,  tend to have larger \civ-BEL blueshift than non-BAL QSOs (e.g., \citealp{Richards2011,Rankine2020}), \civ-BEL blueshift correlates with BAL-trough velocity and width (\citealp{Rankine2020}), and the BAL fraction appears to increase as the increase of \civ-BEL blueshift at fixed \civ-BEL EW (\citealp{Rankine2020}).  We propose that BEL blueshift and BAL could be different manifestations of the same outflow system viewed at different sightlines and/or phases. This can be easily understood because, for a bulk outflow system with different-form QSO winds (e.g., BALs, mini-BALs, or NALs) along different sightlines, BEL blueshift can be detected along most sightlines due to the fact that each outflow unit can absorb as well as emit photons. As a comparison, BALs can be detected only in specific slightlines with sufficient physical conditions that allow their formation. For an individual QSO, if the BAL winds carry more material in the outflow system than other-form QSO winds, then the largest BEL blueshift is likely observed along or close to the BAL sightline (for a demonstration see Fig. 5 from \citealp{Elvis2000}). 

On the other hand, the \oiii\ weakness is known to depend on quasar luminosity (e.g., \citealp{Sulentic2004, Baskin2005, Stern2012}). Previous studies also found that the \oiii\ weakness tends to be associated with \civ-BEL  blueshift based on the investigations of luminous non-BAL quasars (e.g., \citealp{Netzer04,Vietri18}). As a comparison, the BAL composite characterized by a striking \civ-BEL blueshift and weak \oiii\ emission suggests that the \oiii\ weakness is more strongly correlated with the \civ-BEL blueshift than with luminosity, given that our sample have a similar luminosity range to the non-BAL sample of Zuo15. Intriguingly, this is in good agreement  with \citet{Coatman2019}, where they found a much stronger anti-correlation between \civ-BEL blueshift and \oiii\ EW than that between \civ-BEL blueshift and bolometric luminosity from a sample consisting of 213 non-BAL quasars at $2<z<4$. 
Although large \civ-BEL blueshift is most likely caused by nuclear outflows, other possibilities cannot be firmly ruled out either (e.g., \citealp{Gaskell1982}). Unlike the non-BAL sample studies mentioned above,  BALs are unambiguously produced from nuclear outflows. Therefore, our investigations of the relation between \civ-BEL blueshift and \oiii\ emission among BAL QSOs provide complementary and compelling evidence in support of the presence of nuclear outflows affecting the narrow emission-line regions, which is notably characterized by the lack of narrow emission-line profiles as shown in Figure~\ref{nir_composite}. Observationally, the strong anti-correlation between \civ-BEL blueshift and \oiii\ emission (\citealt{Coatman2019}), as well as the co-location of BAL and non-BAL QSOs as a function of \civ-BEL and other physical properties (\citealt{Rankine2020}), together support our argument that BAL and BEL blueshift could be different manifestations of the same nuclear outflow system viewed at different sightlines and/or phases.  In this scenario, it is naturally expected that the presence of substantial nuclear outflows traced either by strong BALs or large BEL blueshifts, can block significant amount of ionizing photons from reaching the NEL region and/or sweep away the NEL gas, and hence lead to weak \oiii\ emission regardless of a specific orientation. This can be tested from individual QSOs in the future.

Last but not least, if these high-$z$ BAL QSOs have biconical NEL regions often seen in low-$z$ AGNs (e.g., \citealp{Liu2013}), our investigations would imply a wide opening-angle scenario for the presence of different-form QSO winds in the same outflow system, which is filled with inhomogeneous absorbers characterized by stratified density/ionization and possibly clumpy structures (e.g., \citealp{Yi19b,Hamann19}). In addition, HiBALs tend to have higher reddening and column densities than non-BALs while LoBALs have even higher reddening and column densities than HiBALs (e.g., \citealp{Reichard03b}), implying that dust could be associated with BAL outflows. 
Conceivably, BAL transitions among LoBAL, HiBAL, and non-BAL states tend to occur within a timescale typically less than five years in the quasar rest frame (e.g., \citealp{Filizak12,McGraw2017,Rogerson2018,Yi19a}). These observational results support an inhomogeneous, dusty outflow system rather than a simple thin-shell outflow system filled with homogeneous absorbers. However, the detailed investigation of such an outflow system requires a significantly large BAL sample based on optical/near-IR spectroscopy to cover both \civ\ and \hb,  dedicated observations for individual quasars with the presence of substantial outflows, and suitable modellings, which is beyond the scope of this work.

\section{Conclusion and future work}  \label{sec:conclusion}

High-redshift BAL QSOs are rare and have been studied sparsely, usually only including individual objects or very small samples.   In this work, we present the largest sample study (to our knowledge) based on 22 high-redshift BAL QSOs  via optical/near-IR spectroscopy. Due to the lack of a similar sample study based on high-$z$ BAL QSOs in the literature, we  compared them with non-BAL QSOs matched in luminosity and/or redshift, mainly focusing on the investigations of nuclear outflows and their effects. Our main results are concluded as follows: 
\begin{enumerate}
\item
We identified 12 low-ionization (Lo) BAL QSOs in this sample, and the fraction ($\sim$54\%) is significantly higher than that from low-z samples (typically $\sim$10\%). This is likely caused by our preferential selection of strong BAL QSOs in constructing the sample (Section \ref{sample_selection}). 

\item 
We construct the composite spectra using seven BAL QSOs with the presence of sub-relativistic outflows,  from which we see the prevalence of large \civ\ emission-line blueshift ($\sim$3100 km s$^{-1}$) and weak \oiii\ emission.  In combination with the same trend found from non-BAL samples, our investigations   provide complementary and compelling evidence that the presence of nuclear outflows in a QSO is indeed capable of affecting its narrow emission-line region (see Section \ref{composite_spec}).  

\item
In the BAL sample, the 3000 \AA\ and 5000 \AA\ continuum luminosities show a strong correlation, consistent with previous non-BAL QSO studies; however, there is no correlation between the \mgii\ and \hb\ lines in FWHM, likely due to the effect of nuclear outflows traced by BALs and \civ\ broad emission-line blueshifts. Together with the striking features shown in the composite constructed from the seven LoBAL QSOs, our investigations offer strong evidence for nuclear outflows influencing the broad emission-line regions  (see Section \ref{non_bal_sample}).

\item
In the distribution of BH mass versus Eddington ratio, our sample shows only a tentative correlation at a marginally significant level compared to strong correlations at highly significant levels from the  three non-BAL comparison samples matched in luminosity and redshift, again, possibly due to an outflow effect in our sample (see Section \ref{comparison_match_z_L}).

\item
Our observational results indicate that these high-redshift BAL QSOs on average do not have a higher Eddington ratio than that from non-BAL QSOs matched in luminosity and/or redshift (see Section~\ref{comparison_z}). 

\item
We propose that BAL and BEL blueshift could be different manifestations of the same outflow system viewed at different sightlines and/or phases (see Section \ref{discussion_section}). 

\end{enumerate}

Our systematic investigations of the similarities and differences between BAL and non-BAL quasars at high redshift, allow us to explore nuclear outflows and their effects from a statistical view. 
We propose that strong BAL and large \civ-BEL blueshift trace the same outflow system, which in turn can explain the striking \civ-BEL blueshift in the BAL sample, as well as the prevalence of large \civ-BEL blueshift and weak \oiii\ emission found from both the BAL and non-BAL samples. 
 BALs are widely accepted as smoking-gun signatures of quasar winds, therefore our observational results provide complementary insight into the different manifestations of ionized outflows and offer strong evidence in support of these outflows affecting the broad and narrow emission-line regions.

We leave the systematic investigations of LoBAL QSOs regarding the optical \feii\ strength, luminosity, Eddington ratio, \civ-BEL blueshift, \oiii\ strength etc, to another dedicated work in combination with two uniform LoBAL samples at $z<1$ \citep{Yi19a} and $1<z<2.5$ \citep{Schulze17}, respectively. As the accumulation of high-$z$ LoBAL QSOs in the future, this may allow us to investigate the redshift evolution in the LoBAL population. However, it could be difficult to obtain a significantly large sample of high-$z$ BAL QSOs in a couple of years due to their rarity as well as the challenge in IR spectroscopy. Our spectroscopic investigations of BAL outflows and their effects based on the optical/near-IR spectroscopy therefore deliver the best statistical results of high-$z$ BAL QSOs to date. In addition, we also plan to investigate individual BAL QSOs showing extreme or peculiar phenomena, aiming to locate  outflow distances, constrain outflow structures and physics, or explore the detailed process of how nuclear outflows influence the large-scale regions, which might be achieved via multi-epoch/wavelength observations and spatially resolved spectroscopy, particularly with the aid of adaptive optics (AO) or future facilities like James Webb Space Telescope (JWST). Such investigations will greatly improve our understanding of quasar winds and their effects in the context of quasar feedback.

\section{Acknowledgements}

We thank Tinggui Wang for stimulating discussion in this work. 
We acknowledge the support of the staff of the Lijiang 2.4m telescope (LJT). Funding for the telescope has been provided by CAS and the People's Government of Yunnan Province. 
This research uses data obtained through the Telescope Access Program (TAP), which has been funded by the National Astronomical Observatories of China, the Chinese Academy of Sciences (the Strategic Priority Research Program "The Emergence of Cosmological Structures" Grant No. XDB09000000), and the Special Fund for Astronomy from the Ministry of Finance. Observations obtained with the Hale Telescope at Palomar Observatory were obtained as part of an agreement between the National Astronomical Observatories, Chinese Academy of Sciences, and the California Institute of Technology. 

W. Yi thanks the financial support from the program of China Scholarships Council (No. 201604910001) for his postdoctoral study at the Pennsylvania State University. W. Yi also thanks the support from the National Science Foundation of China (NSFC-11703076) and the West Light Foundation of the Chinese Academy of Sciences (Y6XB016001). X. Wu thanks the support from the National Key R\&D Program of China (2016YFA0400703) and the National Science Foundation of China (11533001 \& 11721303). This work is also supported by the Joint Research Fund in Astronomy (U1631127) under cooperative agreement between the National Science Foundation of China and Chinese Academy of Sciences.

\bibliographystyle{mnras}
\bibliography{the_entire_lib}

\end{document}